\documentclass[aps,prb,twocolumn,superscriptaddress,nofootinbib,longbibliography]{revtex4-2}
\usepackage[pdftex]{graphicx}
\usepackage[colorlinks=true,citecolor=blue]{hyperref}
\usepackage{graphicx,textcomp,dcolumn,hyperref,upgreek}

\usepackage{physics}

\usepackage{url}

\usepackage{dsfont}

\usepackage{float}

\def\ra{\rightarrow}

\def\>{\rangle}
\def\<{\langle}

\renewcommand{\Re}{\mbox{Re}}
\renewcommand{\Im}{\mbox{Im}}

\def\d{\partial}

\def\h{\hat}

\def\eps{\varepsilon}

\def\0{{(0)}}

\newcommand{\ce}{\text{ce}}
\newcommand{\Sine}{\sin\left(\zeta / \rho\right)}
\newcommand{\Cose}{\cos\left(\zeta / \rho\right)}

\begin{document}

\title{
Electronic transport in bent carbon nanotubes
}
\author{Eric~Kleinherbers}
\email{eric.kleinherbers@uni-due.de}
\affiliation{Faculty of Physics and CENIDE, University of Duisburg-Essen, 47057 Duisburg, Germany}
\affiliation{Department of Physics and Astronomy, University of California, Los Angeles, California 90095, USA}

\author{Thomas~Stegmann}
\email{ stegmann@icf.unam.mx}
\affiliation{Instituto de Ciencias Físicas, Universidad Nacional Autónoma de México, 62210 Cuernavaca, Mexico}

\author{Nikodem~Szpak}
\email{nikodem.szpak@uni-due.de}
\affiliation{Faculty of Physics and CENIDE, University of Duisburg-Essen, 47057 Duisburg, Germany}
               
\date{\today}

\begin{abstract}
We study the electronic transport through uniformly bent carbon nanotubes. For this purpose, we describe the nanotube with the tight-binding model and calculate the local current flow by employing non-equilibrium Green's functions (NEGF) in the Keldysh formalism. In addition, we describe the low-energy excitations using an effective Dirac equation in curved space with a strain-induced pseudo-magnetic field which can be solved analytically for the torus geometry in terms of the Mathieu functions. We obtain a perfect quantitative agreement with the NEGF results.
For nanotubes with an armchair edge, already a weak bending of $1\,\%$ substantially changes the electronic properties. Depending on the valley, the current of the zero mode flows either on the outer or the inner side of the torus and, therefore, can be used as a valley splitter. In contrast, the zigzag nanotubes are largely unaffected by the bending.
Our findings are of importance for nanoelectronic applications of carbon nanotubes and open new possibilities for valleytronics. 
\end{abstract}

\maketitle

%%%%%%%%%%%%%%%%%%%%%%%%%%%%%%%%%%%%%%%%%%%%%%%%%%%%%%%%%%%%%%%%%%%%%%%%%%%%%%%%%%%%%%%%%%%%%%%%%%%%
\section{\label{sec:Introduction} Introduction}
One of the most spectacular facts about graphene is that the low-energy electronic excitations can be, in a good approximation, described by a two-dimensional massless Dirac equation known from the relativistic quantum field theory~\cite{Novoselov+Geim2005, Katsnelson2007, Fialkovsky2012}. A lot of literature has already been devoted to the discussion of possible applications and extensions of this effective picture~\cite{CastroNeto2009review,Katsnelson2020,Torres2020}. Regarding the nanoelectromechanical properties of the material~\cite{gentile_2022, Ortiz2022}, one intriguing observation relates the elastic deformations of the honeycomb lattice with an effective artificial gauge potential that couples to the Dirac field similarly to an electromagnetic vector potential~\cite{manes_2007,vonoppen_2009,vozmediano_2010,wakker_2011,mucha_2012,kitt_2012,neek-amal_2012,CarrilloBastos2014,Naumis2017}. 
At the same time, the Dirac field couples to the induced curvature of the deformed two dimensional surface which offers a unique quantum simulator of the Dirac equation in curved spaces~\cite{dejuan_2007,gonzalez_2010,stegmann_2016,castro-villareal_2017,gallerati_2021}.

\begin{figure}[t]
	\includegraphics[width=8.6cm]{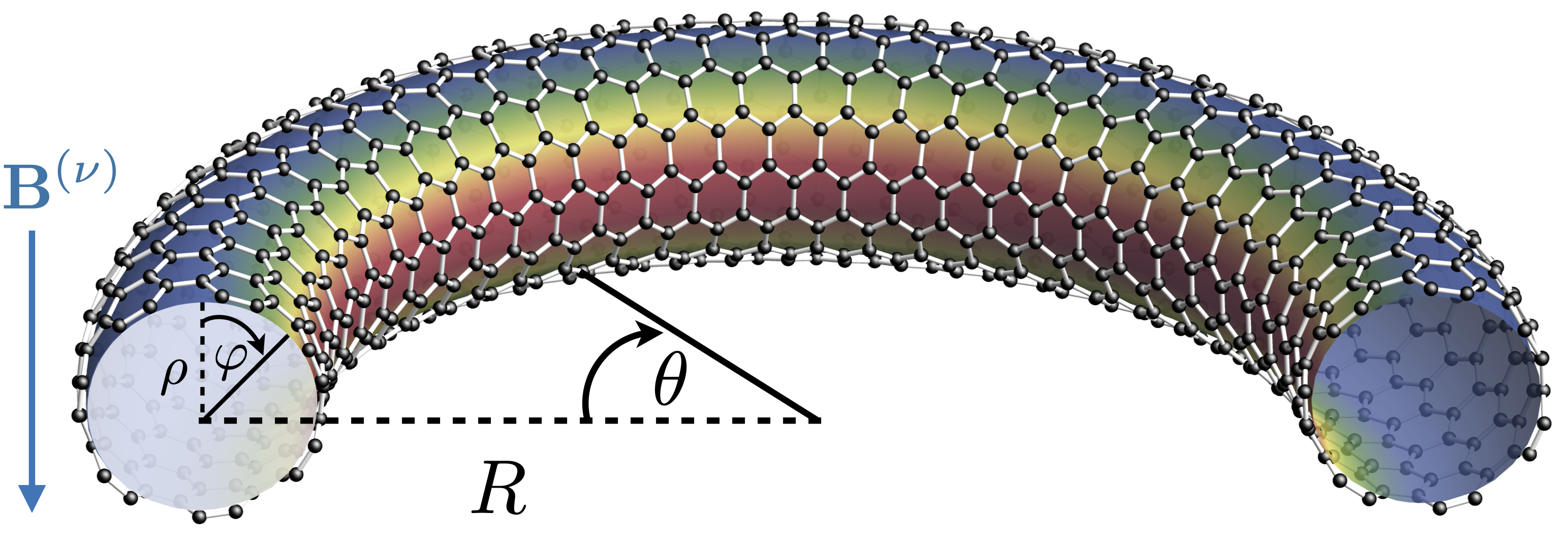}
	\caption{A carbon nanotube with armchair edge uniformly bent into a segment of a torus. The color shading sketches the current flowing in $\theta$-direction carried by electrons of the valley $\nu=+1$ (calculated for a much larger system with parameters of Fig.~\ref{fig:8}). Due to the pseudo-magnetic field $\vb B^{(\nu)}$, effectively perpendicular to the torus, the electrons flow mainly on the inner side (red). In contrast, for $\nu=-1$, they flow on the outer side of the torus (not shown).
	}
	\label{fig:1}
\end{figure}

Here, we focus on the electronic transport in carbon nanotubes (CNT)~\cite{roche_2007,charlier_2007,dubois_2009,laird_2015} which are basically graphene sheets rolled into cylinders. They have been synthesized and studied even before graphene~\cite{iijima_1991, Saito1998, Reich2004}.
Moreover, the nanoelectromechanical properties~\cite{lassagne_2009,eichler_2011, palyi_2012,wang_2016} and, in particular, elastic deformations have repeatedly gained attention~\cite{kane_1997,rochefort_1999,Tombler2000,suzuura_2002,farajian_2003,fa_2004,koskinen_2010,wang_2010,shima_2012,rahman_2017,wu_2019}. 
In this work, we study uniformly bent nanotubes which take the form of a segment of a torus (cf. Fig. \ref{fig:1}). Thus, in contrast to mathematical tori~\cite{ceulemans_2000,Zhang_2005} or tori constructed from defects~\cite{Lizhao_2014}, we study a doughnut-shaped nanotube realized by elastic deformations.
This particular geometry has the advantage that it possesses an interesting strain structure while 
retaining a translational symmetry in the toroidal direction, thus being still tractable analytically. 

The strain varies only in the poloidal direction, being  compressive on the inner side and tensile on the outer side. 
This gives rise to an effective pseudo-magnetic field behaving analogously to an external homogeneous magnetic field oriented orthogonally to the bent nanotube (cf. Fig. \ref{fig:1}). However, the crucial difference is that it couples with different signs to the different valley degrees of freedom. 
This opens the possibilities of interesting valley separation phenomena which we observe and explain below. Such valley splitters are the key devices for a new type of electronics named \textit{valleytronics} which uses the valley degree of freedom of electrons instead of their spin or charge \cite{chaves_2010,Settnes2016, Milovanovic2016,Schaibley2016,osika_2017,CarrilloBastos2018,Sandler2018,Stegmann2018, Ortiz2022, yu_2022}. 
Since carbon nanotubes are considered as a possible building block of future computer chips~\cite{franklin_2012,Shulaker2013, tulevski_2014,Hills2019}, interconnects~\cite{Brand2008, Todri2017} or as nanoelectromechanical sensors~\cite{hierold_2007, Schroeder2018}, understanding the current flow in CNTs is of crucial importance, in particular with respect to the influence of bending and deformation.

This paper is organized as follows. In Sec.~\ref{sec:system}, we introduce the discrete system and calculate within the tight-binding approach the band structure of a bent nanotube with zigzag (ZZ) and armchair (AC) edges. Then, in Sec.~\ref{sec:continuum_model}, we switch to the continuous description and introduce the Dirac equation in curved space with the pseudo-magnetic field and solve it both numerically and analytically on a segment of a torus. In Sec.~\ref{sec:keldysh}, the results are compared with electronic transport calculations using non-equilibrium Green's functions (NEGF) in the Keldysh formalism. Finally, in Sec.~\ref{sec:conclusions}, we conclude our findings.

\section{System} \label{sec:system}
In this work, we consider a carbon nanotube that is bent to a segment of a torus, see Fig.~\ref{fig:1}, which is characterized by the inner radius of the tube $\rho$  and the outer radius $R$.
The ratio of both radii, $\gamma=\frac{\rho}{R}$, is a natural measure of the tube bending. Note that the bending in Fig.~\ref{fig:1} is greatly exaggerated, since we consider here only $\gamma\leq 1\,\%$.
The position on the surface of the torus can be described by the 
toroidal angle $\theta$ and the poloidal angle $\varphi$.
Due to the strain, the distances of the carbon atoms in the toroidal direction are shortened on the inner side of the torus ($\varphi=\pi/2$) while they are elongated on the outer side ($\varphi=-\pi/2$).

Here, we describe the $\pi$-electron system by the Schrödinger equation $H\ket{\Psi}=E\ket{\Psi}$ with the simple tight-binding Hamiltonian 
\begin{align}
H&=-\sum_{\langle\vb n,\vb m\rangle} t_{\vb n \vb m} \left(\dyad{\vb{m}}{\vb n} + \dyad{\vb{n}}{\vb m}\right),\label{eq:tightbinding}\\
&=\sum_{n,k} \epsilon_n(k) \dyad{n,k},
\end{align}
where the summation runs only over nearest neighbors $\langle\vb n, \vb m\rangle$.
Since the honeycomb lattice of graphene consists of two interconnected triangular lattices, $\cal A$ and $\cal B$,
only the $\pi$ orbitals at different sublattices, $\vb{n}\in{\cal A}$ and  
$\vb{m}\in{\cal B}$, 
are coupled. The tunneling amplitudes $t_{\vb{n}\vb{m}}$  depend on the poloidal positions of the atoms $\vb{n}$ and $\vb{m}$. 
In the second line, we formally write down the diagonalized Hamiltonian, where 
$\ket{n,k}$ are the Bloch states  with energy $\epsilon_n(k)$ in band $n$ with wave number $k$ (in the $\theta$-direction). 

In pristine graphene, the distances between the neighboring carbon atoms are all equal, $d_0=0.142\,\text{nm}$, and hence the tunneling amplitudes are identical, $t_{\vb{n} \vb{m}}=t_0=2.8\,\text{eV}$.  However, when strain is applied they get modified according to the empirical formula 
\begin{align}\label{eq:tunnelmod}
t_{\vb n \vb m}=t_0 e^{-\beta\left( \vert \vb n- \vb m\vert/d_0 -1\right)},
\end{align}
where $\beta\approx3.37$ is the material specific Grüneisen parameter~\cite{Carrillo-Bastos2016, Pereira2009, Ribeiro2009}. 
In consequence, on the torus, the tunnel couplings in  toroidal direction
become slightly larger on the inner side than on the outer side. As we restrict ourselves to bendings $\gamma \le 1\%$, the resulting maximum relative change of the tunneling amplitudes is of the order of $\delta t/t_0 \sim \gamma \beta \sim 3.37\%$.

In this paper, we consider large tube diameters ($\gg1\,\text{nm}$) for which we can neglect the  misorientation of $\pi$ orbitals~\cite{kane_1997,dubois_2009}.
Its effect can be estimated via the Slater-Koster formula, $\delta t/t_0 = 1 - \cos(\delta\phi)$, where $\delta \phi \sim d_0/\rho$ is the relative angle between the locally tilted (orthogonal to the nanotube's surface) $\pi$-orbitals \cite{Kleiner+Egger1, Kleiner+Egger2}.
For diameters $\gg$ 1nm we obtain  $\delta t/t_0 \ll 1\%$, i.e., the effect of misorientation of $\pi$ orbitals is negligible compared to strain. 
By a similar reasoning, we can neglect atomic~\cite{gmitra2009band} and curvature-enhanced spin-orbit coupling~\cite{huertas_2006,jeong_2009,izumida_2009,klinovaja_2011}, since the latter scales with $\sim \rho^{-1}$.

\begin{figure}[t]
	\includegraphics[width=.49\textwidth]{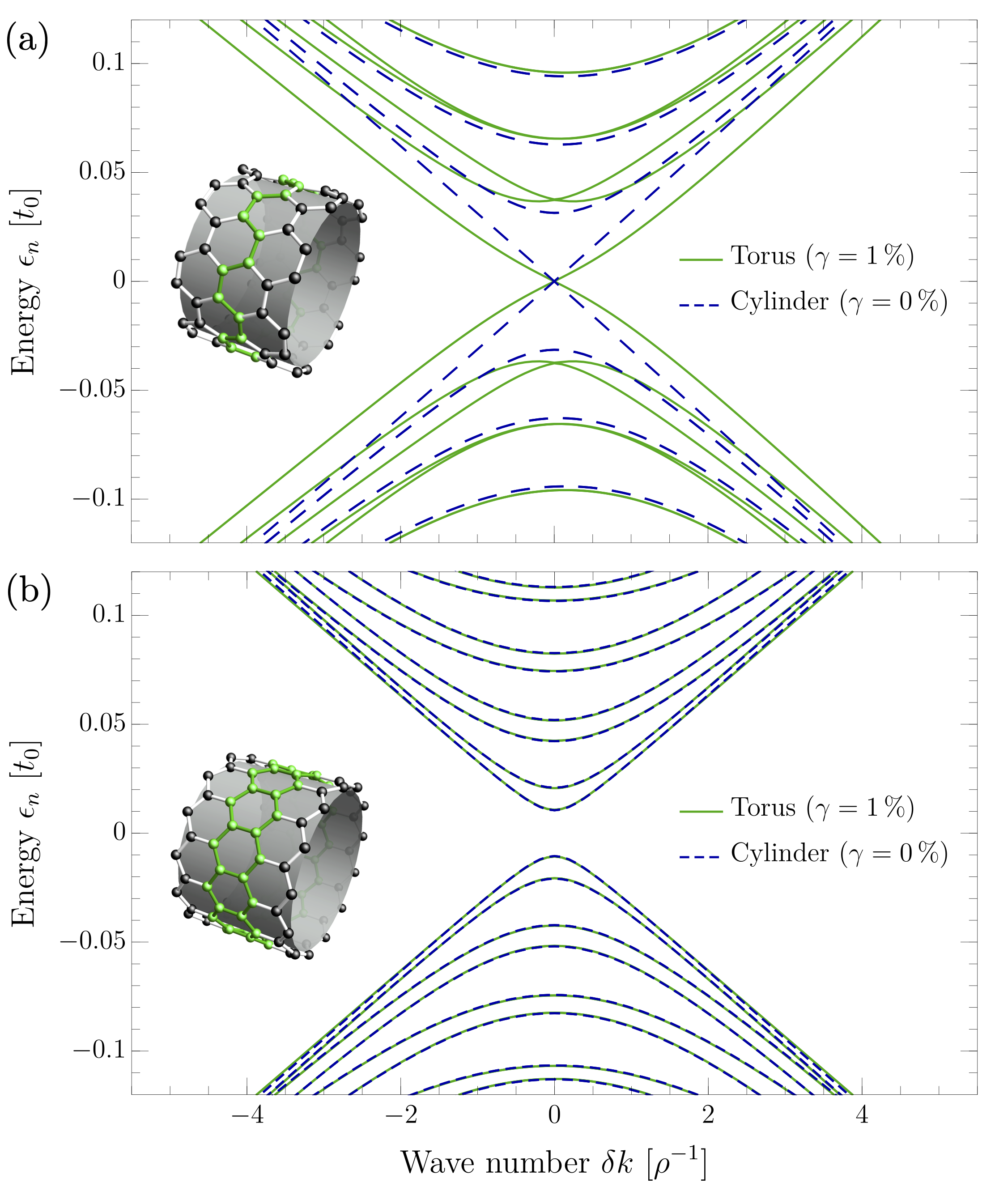}
	\caption{Band structure of a torus-shaped (a) armchair and (b) zigzag nanotube as a function of the wave number $\delta k$ relative to the Dirac point, where the bending is $\gamma=1\,\%$ (green lines). For comparison, the respective band structure for a straight nanotube is shown (blue dashed lines). The inset indicates the unit ring (green) used for the band structure calculation. We used 400 atoms for (a) and 692 atoms for (b) in each ring. The radius is given by $\rho\approx47.7\,d_0$.}
	\label{fig:2}
\end{figure}

As a first step, we discuss the electronic excitation energies of the system. 
For straight nanotubes, the periodic boundary condition along the circumference gives rise to a band structure composed of vertical cuts through the characteristic Dirac cone of planar graphene, see Fig.~\ref{fig:2}. Depending on the chirality, these cuts can either pass directly through the Dirac points or beside it. In the former case, one obtains a gapless, linear dispersion relation and in the latter case, one obtains a band gap~\cite{laird_2015}. Here, we discuss only two special chiralities, where the edge of the nanotube is either in the armchair, Fig.~\ref{fig:2}(a), or the zigzag direction, Fig.~\ref{fig:2}(b). In the armchair case, the two inequivalent Dirac points are folded back to $\pm 2\pi/(3\sqrt{3}d_0)$, while for the zigzag case they merge at the $\Gamma$-point~\cite{dubois_2009}.

For bent nanotubes, we calculate the band structure by identifying the unit ring which is periodically repeated in toroidal direction, see the green atoms in the insets of Fig.~\ref{fig:2}(a)-(b). 
Thus, we assume infinite nanotubes where the rings occur with (super) lattice constants $a_{\text{AC}}=\sqrt{3}d_0$ and  $a_{\text{ZZ}}=3d_0$ for the armchair and zigzag case, respectively. Then, by using Bloch‘s theorem, we determine the band structure, $\epsilon_n(k)$, shown in Fig.~\ref{fig:2}.
We remark that for finite nanotubes, we expect for the zigzag case additional localized edge states~\cite{nakada_1996,sasaki_2005}.

In Fig.~\ref{fig:2}(a), the bands $\epsilon_n(\delta k)$ (green lines) are shown for an armchair nanotube, where $\delta k$ is the wavenumber relative to the Dirac point. Already a small bending of $\gamma=1\,\%$ noticeably modifies the band structure compared to a straight nanotube (blue dashed lines). Although the Fermi velocity changes, the spectrum is still gapless. 
In contrast, for the bands of a zigzag nanotube, the same bending of $\gamma=1\,\%$  has almost no influence, see Fig.~\ref{fig:2}(b).

In the following, we study the bent nanotube in the continuum limit using an effective Dirac equation in curved space and we will understand why the influence of bending is much more severe for armchair nanotubes than for zigzag nanotubes.

%%%%%%%%%%%%%%%%%%%%%%%%%%%%%%%%%%%%%%%%%%%%%%%%%%%%%%%%%%%%%%%%%%%%%%%%%%%%%%%%%%%%%%%%%%%%%%%%%%%%
\section{\label{sec:continuum_model} Continuum model}
%%%%%%%%%%%%%%%%%%%%%%%%%%%%%%
In the low-energy expansion around the Dirac cones, the tight-binding model with smooth and small deformations~\eqref{eq:tightbinding} can be reduced to the continuous Dirac equation  in curved space~\cite{dejuan_2007,vozmediano_2010,stegmann_2016} for the spinor $\underline{\psi}=(\psi_+,\psi_-)$, where the sign $\pm$ originates from the sublattice index and is also referred to as \textit{pseudo} spin.
The Dirac Hamiltonian ${\cal H}_D$ for the valley $\nu=\pm 1$ takes the form ($\hbar=1$)
\begin{align}\label{eq:dirac}
 {\cal H}_D \underline{\psi}&= \vb{v}_F(\vb x) \cdot \left[\vb p - \vb K^{(\nu)}(\vb x) -i\,\boldsymbol{\Omega}(\vb x)\right] \underline{\psi} \nonumber \\
  &= -i v_F \sigma^a e_a^{~j}(\vb x)  \left[\d_j -i K_j^{(\nu)}(\vb x)  + \Omega_j(\vb x)\right] \underline{\psi},
\end{align}
where  $\vb{v}_F(\vb x)$  is the position dependent matrix valued Fermi velocity, $\vb p$ is the momentum operator,  $\vb K^{(\nu)}(\vb x)$ is the shifted position of the Dirac cone due to the deformation, and $\boldsymbol{\Omega}(\vb x)$ is the spin connection~\cite{vozmediano_2010}. In the second line, we use from henceforth the Einstein summation convention and write the Fermi velocity with its contravariant components ${v}^j_F(\vb x)=v_F \sigma^a e_a^{~j}(\vb x)$, where  $v_F=3t_0d_0/2$ is the flat space value, $\sigma^a$ are the Pauli matrices, and  $e_a^{~j}(\vb x)$ are the components of the frame field $\vb e_a(\vb x)$ (\textit{zweibein}) with $a=1,2$. By inserting the momentum operator $p_j=-i \d_j$, we can identify the proper covariant derivative $\d_j+\Omega_j(\vb x)$ for spinors in a local frame. 
Finally, the position of the Dirac cone $K_j^{(\nu)}(\vb x)$ depends now on $\vb x$ and couples to the spinor wave function similarly to an electromagnetic vector potential. Its curl can create a pseudo-magnetic field $B^{(\nu)}$ --- \textit{pseudo} because it maintains time reversal symmetry by pointing in opposite directions in the two valleys $\nu=+1$ and $\nu=-1$.
%%%%%%%%%%%%%%%%%%%%%%%%%%%%%%

In the following, the effective Dirac equation~\eqref{eq:dirac} will be solved analytically for the geometry of a uniformly bent nanotube forming a segment of a torus, see Fig.~\ref{fig:1}. We assume large radii $2\pi\rho\gg d_0$ where the continuum model is expected to give a valid description of the system.
We emphasize that the torus, in contrast to the cylinder, has a real, intrinsic curvature. 
Above, we already introduced the toroidal angular coordinate $\theta$ and the poloidal angular coordinate $\varphi$, see Fig.~\ref{fig:1}. 
For dimensional reasons, however, it is convenient to use the coordinates $\vb{x}=(x^1, x^2) = (\xi, \zeta)$ defined by $\xi=R\,\theta$, $\zeta=\rho\,\varphi$ and having the unit of length.

%%%%%%%%%%%%%%%%%%%%%%%%%%%%%%%%%%%%%%%%%%%%%%%%%%%%%%%%%%
\subsection{Strain tensor}
%%%%%%%%%%%%%%%%%%%%%%%%%%%%%%%%%%%%%%%%%%%%%%%%%%%%%%%%%
For the torus--like deformation we get a tensile strain on the outer side and a compressive strain on the inner torus side. 
The strain tensor of the 2D torus surface
\begin{equation}
\hat{ \boldsymbol{\eps}}= \begin{pmatrix}
               \eps_{\xi\xi} & \eps_{\xi\zeta} \\
               \eps_{\zeta\xi} & \eps_{\zeta\zeta}
            \end{pmatrix}
\end{equation}
has only one non-zero entry
\begin{align}\label{eq:strain}
  \eps_{\xi\xi}=-\gamma\Sine,
\end{align}
where we consider the effects of strain only up to the first order of $\gamma$. Since we restrict ourselves to $\gamma\le 1\%$, we assume that the strain is adequately described by Eq.~\eqref{eq:strain} and we do not need to perform any atomic relaxation calculations~\cite{seon-myeong_2010}.

%%%%%%%%%%%%%%%%%%%%%%%%%%%%%%%%%%%%%%%%%%%%%%%%%%%%%%%%%%
\subsection{Pseudo magnetic field}
%%%%%%%%%%%%%%%%%%%%%%%%%%%%%%%%%%%%%%%%%%%%%%%%%%%%%%%%%
The deformation shifts the Dirac points from pristine graphene $\vb K_0^{(\nu)}$ to $\vb K^{(\nu)}(\vb x)=\vb K_0^{(\nu)}+\vb A^{(\nu)}(\vb x)$,
where the pseudo-magnetic vector potential $\vb A^{(\nu)}(\vb x)$ can be obtained directly from the strain $\hat{ \boldsymbol{\eps}}$~\cite{stegmann_2016}
and depends on the nanotube orientation (chirality). %
For the AC--CNT, we find
\begin{align}
  \vb K^{(\nu)}&=\vb K_0^{(\nu)}+\nu\, \frac{\beta}{2}
  \begin{pmatrix}
 {\eps}_{\zeta\zeta}-  {\eps}_{\xi\xi} \\
    -2{\eps}_{\xi\zeta} 
  \end{pmatrix}\\
  &=\nu \left[\begin{pmatrix}
    \frac{4\pi}{3\sqrt{3} d_0} \\ 0
  \end{pmatrix}+ \begin{pmatrix}
  \frac{\beta{\gamma}}{2 d_0}\Sine \\ 0
  \end{pmatrix} \right].
  \label{a_vekac}
\end{align}
For the ZZ--CNT, we find
\begin{align}
  \vb K^{(\nu)}&=\vb K_0^{(\nu)}+\nu\,\frac{\beta}{2}\begin{pmatrix}
    -2{\eps}_{\xi\zeta}  \\
    {\eps}_{\zeta\zeta} -{\eps}_{\xi\xi}
  \end{pmatrix}\\
 & =\nu \left[\begin{pmatrix}
    0 \\ \frac{4\pi}{3\sqrt{3}d_0} 
  \end{pmatrix}+ \begin{pmatrix}
    0 \\  \frac{\beta{\gamma}}{2d_0} \Sine 
  \end{pmatrix}\right].
\end{align} 
Notice that the pseudo-magnetic vector potential $\vb A^{(\nu)}$ points in both cases in the ZZ--direction. 
In consequence, the pseudo-magnetic field
$B^{(\nu)} = \nabla \cross \vb{{A}}^{(\nu)}=\epsilon^{ij}\partial_i {A}^{(\nu)}_j$
vanishes in the ZZ case whereas in the AC case it is 
\begin{align}
  B^{(\nu)}=-\nu\,\frac{\beta{\gamma}}{2\rho \, d_0}\Cose=-\nu B_0 \Cose,\label{eq:pseudo_mag}
\end{align}
where we introduced $B_0=\beta \gamma/(2\rho d_0)$.
Embedded in three dimensions, this pseudo-magnetic field takes a surprisingly simple form of a homogeneous field $\vb{B}^{(\nu)}$ pointing parallel to the direction perpendicular to the plane of the torus, see Fig.~\ref{fig:1}. Then, by projecting it onto the surface one obtains again Eq.~\eqref{eq:pseudo_mag}.
Compared to real magnetic fields, these pseudo-magnetic fields are quite strong~\cite{levy_2010,kim_2011}. For the parameters of Fig.~\ref{fig:2}, we find $\hbar B_0/e=11.5\,\text{T}$.
% 

%%%%%%%%%%%%%%%%%%%%%%%%%%%%%%%%%%%%%%%%%%%%%%%%%%%%%%%%%%
\subsection{Fermi velocity (effective frame and metric)}
%%%%%%%%%%%%%%%%%%%%%%%%%%%%%%%%%%%%%%%%%%%%%%%%%%%%%%%%%
The position dependent Fermi velocity $\vb{v}_F(\vb x)$ can be obtained from the orthonormal frame field $\vb{e}_a(\vb{x})$.
For the flat space (corresponding to the straight nanotube) it has the form
\begin{align}
  \vb{e}_1^{(0)}=
  \begin{pmatrix}
    1 \\ 0 
  \end{pmatrix}, \quad
  \vb{e}_2^{(0)}=
  \begin{pmatrix}
    0 \\ 1
  \end{pmatrix},
\end{align}
where $\vb{e}_1^{(0)}$ points in the axial direction and $\vb{e}_2^{(0)}$ points in the azimuthal direction of the nanotube.
For the deformed nanotube, the frame field is modified by the effective strain tensor $\beta \hat{ \boldsymbol{\eps}}$ in leading order of $\gamma$
according to~\cite{dejuan_2012,stegmann_2016}
\begin{align}
  \vb{e}_a(\vb x) &= \left[\mathds{1}-\beta\,{ \hat{ \boldsymbol{\eps}}}\right]\vb{e}_a^{(0)}, 
\end{align}
which gives
\begin{align}
  \vb{{e}}_1(\vb x) &=
  \begin{pmatrix}
    1+\beta \gamma\Sine \\ 0 
  \end{pmatrix}, \\
  \vb{{e}}_2(\vb x) &=
  \begin{pmatrix}
    0 \\ 1
  \end{pmatrix}.
\end{align}
We use the notation $e_a^{~i}$ with $a=1,2$ and $i=\xi,\zeta$ for the components of the frame.
The frame is then orthogonal,
\begin{align}
e_a^{~i}e_b^{~j}g_{ij}=\delta_{ab},
\end{align}
with respect to the effective metric
\begin{align}
  \h g=\mathds{1}+2\beta\,{ \hat{ \boldsymbol{\eps}}}\approx 
  \begin{pmatrix}
    \left[1-\beta \gamma \Sine\right]^{2} & 0 \\
    0 & 1
  \end{pmatrix}.
\end{align}
Without the parameter $\beta$, this is the metric of the torus. Thus, the Dirac equation couples to an \textit{effective} curved geometry where the deformations are enhanced by the material specific Grüneisen parameter $\beta > 1$.

%%%%%%%%%%%%%%%%%%%%%%%%%%%%%%%%%%%%%%%%%%%%%%%%%%%%%%%%%
\subsection{Spin connection}
%%%%%%%%%%%%%%%%%%%%%%%%%%%%%%%%%%%%%%%%%%%%%%%%%%%%%%%%%
The spin connection $\boldsymbol{\Omega}(\vb x)$ can be completely removed from the Dirac equation~\eqref{eq:dirac} (see also Ref.~\cite{kozlovsky_2020}) by the scaling transformation $\underline{\Phi}=e^S\underline{\psi}$ with $S=\log \left( \sqrt[4]{\det \h g}\right)$ (cf. Appendix~\ref{app:spin_connection} for details). 
Then, we obtain
\begin{align}\label{eq:dirac_nospinconnection}
H_D\underline{\Phi}(\vb x)=E \underline{\Phi}(\vb x)
\end{align}
with the rescaled Hamiltonian
\begin{align}
H_D=e^S  {\cal H}_D e^{-S}=-iv_F \sigma^a {e}_a^{~j}(\vb x)\left[\partial_j -i {K}_j^{(\nu)}(\vb x)\right].
\end{align}
%%%%%%%%%%%%%%%%%%%%%%%%%%%%%%%%%%%%%%%%%%%%%%%%%%%%%%%%%%
This transformation also changes the normalization condition for $\underline{\Phi}$.
For the original wave function $\underline{\psi}$, we normalize in curved space via
\begin{align} 
  1 \stackrel{!}{=} \int \mathrm{d^2x}\,\sqrt{\det \h g }\,\underline{\psi}^\dagger \cdot\underline{\psi}  
  = \int \mathrm{d^2x}\,\underline{\Phi}^\dagger\cdot\underline{\Phi}, \label{eq:dirac_norm}
\end{align}
where $ \mathrm{d^2x}\,\sqrt{\det \h g }$ is the integral measure~\cite{stegmann_2016}.
Interestingly, in the second step this measure is completely removed by $\underline{\Phi}=e^S\underline{\psi}$. This means the transformed wave function $\underline{\Phi}$ is the natural object to compare with the discrete solution $\Phi_{\vb n}=\braket{\vb n}{\Psi}$ and $\Phi_{\vb m}=\braket{\vb m}{\Psi}$ of the tight binding model for which the above normalization condition~\eqref{eq:dirac_norm} becomes
\begin{align}
  \sum_{\vb{n}\in \mathcal{A}} \Phi_{\vb{n}}^* \Phi_{\vb{n}}+\sum_{\vb{m}\in \mathcal{B}}\Phi_{\vb{m}}^*\Phi_{\vb{m}}=1, 
\end{align}
by using the lattice discretization $\int \mathrm{d^2x}\ra \sum_{\vb{n}/\vb{m}}$ and $\Phi_\pm(\vb x)\ra\Phi_{\vb{n}/\vb{m}}$ for $\vb{n}\in \mathcal{A}$ and $\vb{m}\in\mathcal{B}$.

%%%%%%%%%%%%%%%%%%%%%%%%%%%%%%%%%%%%%%%%%%%%%%%%%%%%%%%%%
\subsection{Solutions of the Dirac equation}
%%%%%%%%%%%%%%%%%%%%%%%%%%%%%%%%%%%%%%%%%%%%%%%%%%%%%%%%%
In order to find solutions of the Dirac equation, we make use of the translational invariance in the toroidal direction and use the ansatz
\begin{align}
  \underline{\Phi}(\vb x)=e^{i k \xi}\underline{\chi}(\zeta)=e^{ik\xi}
  \begin{pmatrix}
    \chi_+(\zeta) \\ \chi_-(\zeta)
  \end{pmatrix}.
  \label{a_separation}
\end{align}
In the following, we separately discuss the armchair and zigzag case.
%%%%%%%%%%%%%%%%%%%%%%%%%%%%%%%%%%%%%%%%%%%%%%%%%%%%%%%%%
\begin{figure}[t]
	\includegraphics[width=.49\textwidth]{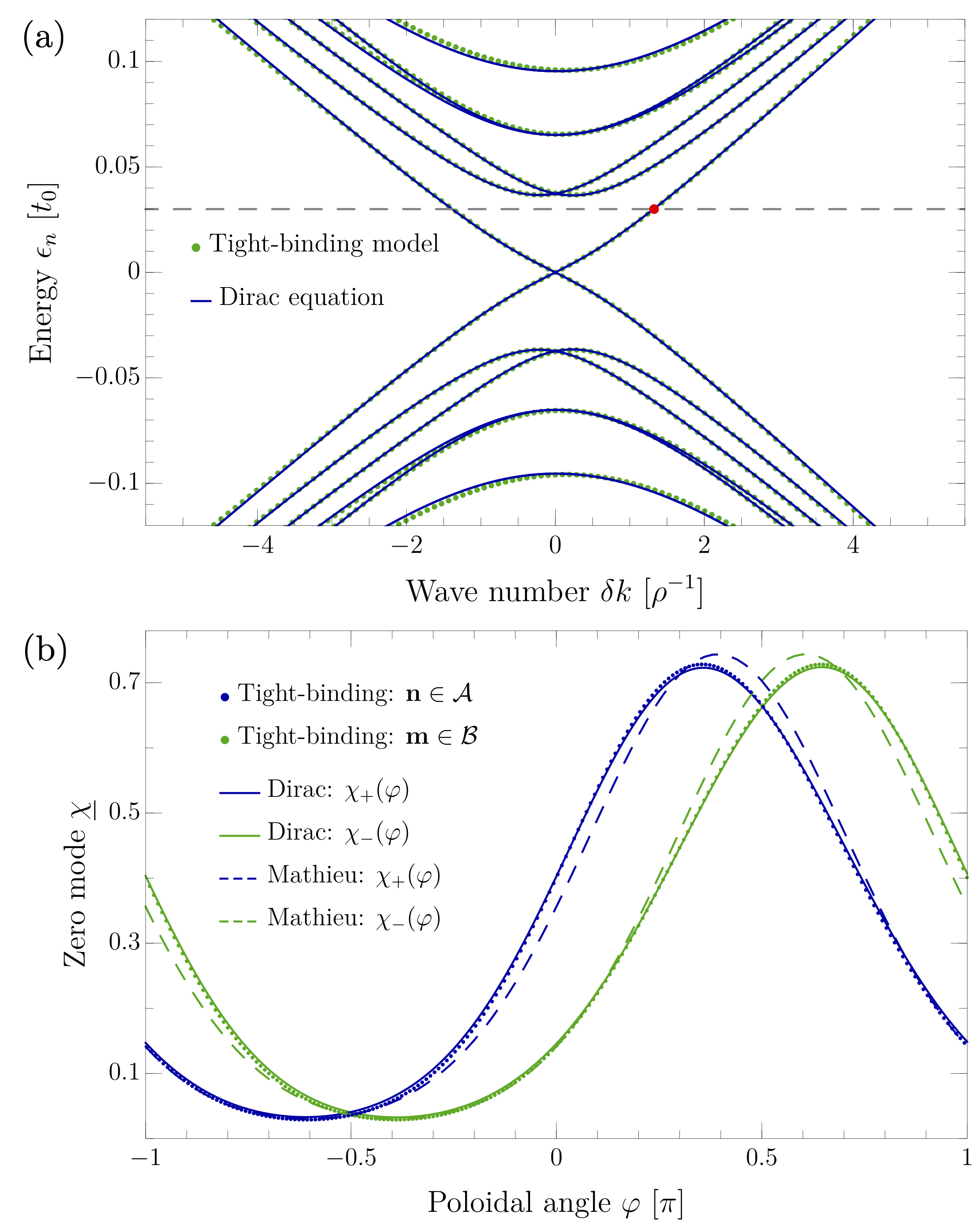}
	\caption{(a) Band structure of a bent  armchair nanotube as a function of the wave number $\delta k$ relative to the Dirac point ($\nu=+1$) with $\gamma=1\,\%$ and $\rho\approx47.7\,d_0$. Numerically obtained solutions from the tight-binding model (green dots) and the Dirac equation (blue line) are compared. (b) The zero mode for sublattice $\cal A$ and $\cal B$ is numerically obtained from the tight-binding model (blue and green dots) and from the Dirac equation (blue and green solid line). In addition, approximated analytical solutions  in form of Mathieu functions are shown (blue and green dashed lines).  The wavenumber is $\delta k=1.33\,\rho^{-1}$ and the energy is $\epsilon_0=0.03\,t_0$ which is indicated by the red dot in (a).}\label{fig:3}
\end{figure}
\subsubsection{Armchair edge}
%%%%%%%%%%%%%%%%%%%%%%%%%%%%%%%%%%%%%%%%%%%%%%%%%%%%%%%%%
Using the product ansatz from above and the convention $(\sigma^1,\sigma^2)=(\nu \,\sigma_x,\sigma_y)$, we obtain from Eq.~\eqref{eq:dirac_nospinconnection} in the leading order in $\gamma$ the Dirac Hamiltonian
\begin{align}
H_D=v_F\left[ -i\sigma_y \d_\zeta +\sigma_x \kappa(\zeta)\right], \label{eq:dirac_armchair}
\end{align}
where we defined
\begin{align}
\kappa(\zeta)=\nu\, \delta k+\left( \nu\,\delta k-\frac{1}{2d_0}\right)\beta \gamma \Sine
\end{align}
with $\delta k=k-K_{0,\xi}^{(\nu)}$.\footnote{Note that switching valleys, $\nu\leftrightarrow -\nu$, and changing the wave numbers according to $\delta k\leftrightarrow-\delta k$ leaves the Hamiltonian unchanged which corresponds to the inversion symmetry of the system.}
As a consistency check of the continuous model, we solve for the eigenvalues and eigenmodes numerically (solid lines) and compare them with the results of the tight-binding model (dots), see Fig.~\ref{fig:3}.
The agreement of the spectrum $\epsilon_n(\delta k)$ for an armchair nanotube in Fig.~\ref{fig:3}(a) (same spectrum as in Fig.~\ref{fig:2}(a)) is almost perfect. Only for higher energies, deviations become visible. Moreover, in Fig.~\ref{fig:3}(b), the lowest-energy eigenstate or \textit{the zero mode} for a given wavenumber $\delta k$ is also surprisingly well described by the continuous model.

Now, in order to find analytical solutions, we first evaluate the squared Hamiltonian and find after some algebra
\begin{align}
H_D^2&=v_F^2 \left[ -i \sigma_y \d_\zeta +\sigma_x \kappa(\zeta)\right]^2 \\
&=v_F^2 \left[- \d^2_\zeta + \kappa^2(\zeta)-\sigma_z \d_\zeta \kappa(\zeta) \right].
\end{align}
Since $H_D^2$ is diagonal, we can solve the eigenvalue problem, $H_D^2 \underline{\chi} =E^2 \underline{\chi}$, by means of the decoupled equations
\begin{align}
\partial^2_\zeta \chi_\pm + \left[(E/v_F)^2-\kappa^2(\zeta)\pm\partial_\zeta \kappa(\zeta)  \right] \chi_\pm&=0.\label{a_entkoppelt+} 
\end{align}
Next, we expand the bracket in the first order in $\gamma$ and find\footnote{Note that the approximation has to be taken with a grain of salt when $\gamma \sim \delta k\, d_0$.}
\begin{align}
(E/v_F)^2-\kappa^2(\zeta)\pm\partial_\zeta \kappa(\zeta)  \approx \frac{a}{4\rho^2} \mp \frac{q}{2\rho^2} \cos(\zeta/\rho \mp \varphi_0),\label{eq:mathieu_approx}
\end{align}
where we have introduced the parameters
\begin{align}
\varphi_0&=\nu \arctan(2\,\delta k\,\rho),\\
a&=\left(\frac{2\rho}{v_F}\right)^2\left[E^2 - (v_F \,\delta k)^2\right],\\
q&=-2{\beta\gamma}\sqrt{1+(2\,\delta k\,\rho)^2}\left(\nu\, \delta k-\frac{1}{2d_0} \right)\rho.
\end{align}
Switching back to the poloidal angle $\varphi=\zeta/\rho$, we obtain
\begin{align}
  \partial^2_\varphi \chi_\pm(\varphi) + \frac{1}{4}\left[a\mp2q \cos(\varphi \pm \varphi_0) \right]\chi_\pm(\varphi) = 0, \label{a_mathieu+-}
\end{align}
which has the form of the \textit{Mathieu} differential equation \cite{abramowitz_1988}.
With the periodic boundary condition $\underline{\chi}(\varphi+2\pi)=\underline{\chi}(\varphi)$, the parameter $a$ becomes restricted to discrete characteristic values $a_{2m}(q)$ and $b_{2n}(q)$.\footnote{The characteristic parameters of the Mathieu equation are available, e.g., in Mathematica by MathieuCharacteristicA[2$m$,$q$] and MathieuCharacteristicB[2$n$,$q$].} 
For $a=a_{2m}$ with  $m=0,1,2,\ldots$, the solutions are given by the even cosine-like functions 
\begin{align}\label{eq:mathieu_cos}
\chi_{\pm}(\varphi)\sim\text{ce}\left(a_{2m},\pm q,\frac{\varphi\pm\varphi_0}{2}\right).
\end{align}
For $a=b_{2n}$ with $n=1,2,\ldots$, the solutions are odd sine-like functions
\begin{align}\label{eq:mathieu_sin}
\chi_{\pm}(\varphi)\sim\text{se}\left(b_{2n},\pm q,\frac{\varphi\pm\varphi_0}{2}\right).
\end{align}
Henceforth, we will only be interested in the zero mode $\chi_{\pm}(\varphi)\sim \text{ce}(a_{0},\pm q,\frac{\varphi\pm\varphi_0}{2})$ which has the lowest positive energy eigenvalue and becomes constant in the limit of vanishing curvature, $\gamma\rightarrow 0$. In Fig.~\ref{fig:3}, this solution (dashed lines) is compared both to the exact eigenmode from the tight-binding  model  (dots) and to the numerical solution of the Dirac equation (solid lines). Deviations originate in the approximation in Eq.~\eqref{eq:mathieu_approx}.

%%%%%%%%%%%%%%%%%%%%%%%%%%%%%%%%%%%%%%%%%%%%%%%%%%%%%%%%%

\subsubsection{Zigzag edge}
%%%%%%%%%%%%%%%%%%%%%%%%%%%%%%%%%%%%%%%%%%%%%%%%%%%%%%%%%
For the zigzag edge, we use the convention $(\sigma^1,\sigma^2)=(\sigma_y,\nu\, \sigma_x)$ and obtain a similar Dirac Hamiltonian
\begin{align}
H_D=v_F\left[ -i\nu \sigma_x \left(\d_\zeta-iK_\zeta^{(\nu)} \right) +\sigma_y \kappa(\zeta)\right], \label{eq:dirac_zigzag}
\end{align}
with
\begin{align}
\kappa(\zeta)= k \left[ 1+\beta \gamma \Sine \right].
\end{align}
Since in this case  the pseudo-magnetic field vanishes, $B^{(\nu)}=0$, the pseudo-magnetic vector potential $K_\zeta^{(\nu)}$ can be gauged away by using
\begin{align}
\underline{{\chi}}= \text{exp}\left[i\int\limits_0^\zeta\mathrm{d}\zeta^\prime K_\zeta^{(\nu)}(\zeta^\prime) \right]\underline{\tilde\chi}.
\end{align} 
Thus, compared to the Dirac equation of the armchair case, where the leading order correction ${\cal O}(\gamma)$ comes from the pseudo-magnetic field, here, we only have the effect of the curvature ${\cal O}(\gamma k)$ which is additionally suppressed by the wavenumber $k\ll 1$. As a consequence, the bending $\gamma$ has a much smaller influence on the electronic properties than in the armchair case, cf. the spectrum Fig.~\ref{fig:2}.
Analogously to the armchair case, we again obtain a Mathieu equation
\begin{align}
 \partial_\varphi^2\tilde\chi_\pm(\varphi) + \frac{1}{4}\left[a\mp2q \cos(\varphi \pm \varphi_0) \right]\tilde\chi_\pm(\varphi) = 0,\label{a_zz_mathieu+-}
\end{align}
with the parameters
\begin{align}
\varphi_0&=- \arctan(2 k\rho),\\
a&=\left(\frac{2\rho}{v_F}\right)^2\left[E^2 - (v_F k)^2\right],\\
q&=2{\beta\gamma}\sqrt{1+(2 k \rho)^2}\, k \rho.
\end{align}
As expected, the parameters are now independent of the valley $\nu=\pm1$ since $B^{(\nu)}=0$. However, it is important to note that the gauge transformation changes the periodic boundary condition to~\cite{kane_1997}
\begin{align}
e^{i k_\varphi 2\pi }\underline{\tilde \chi}(\varphi+2\pi)= \underline{\tilde \chi}(\varphi),
\end{align}
where $k_\varphi=(K^{(\nu)}_\zeta \rho)~\text{mod}~1$  describes the distance between the quantized momentum in the $\varphi$ direction from the Dirac point. Using $\rho=\sqrt{3} N/(2\pi)$ with $N$ being the number of carbon rings along the nanotube edge, we find $k_\varphi= 2N/3~\text{mod}~1$ with the only nonequivalent values  $k_\varphi=0,1/3,2/3$. Since the cases $k_\varphi \neq0$ lead to complicated analytical solutions we focus in the following on the gapless case ($k_\varphi=0$) for $N=3 \,m$ with $m=1,2,\ldots$ and find periodic solutions of the same type as in Eq.~\eqref{eq:mathieu_cos} and Eq.~\eqref{eq:mathieu_sin}.

\begin{figure}[t!]
	\includegraphics[width=.49\textwidth]{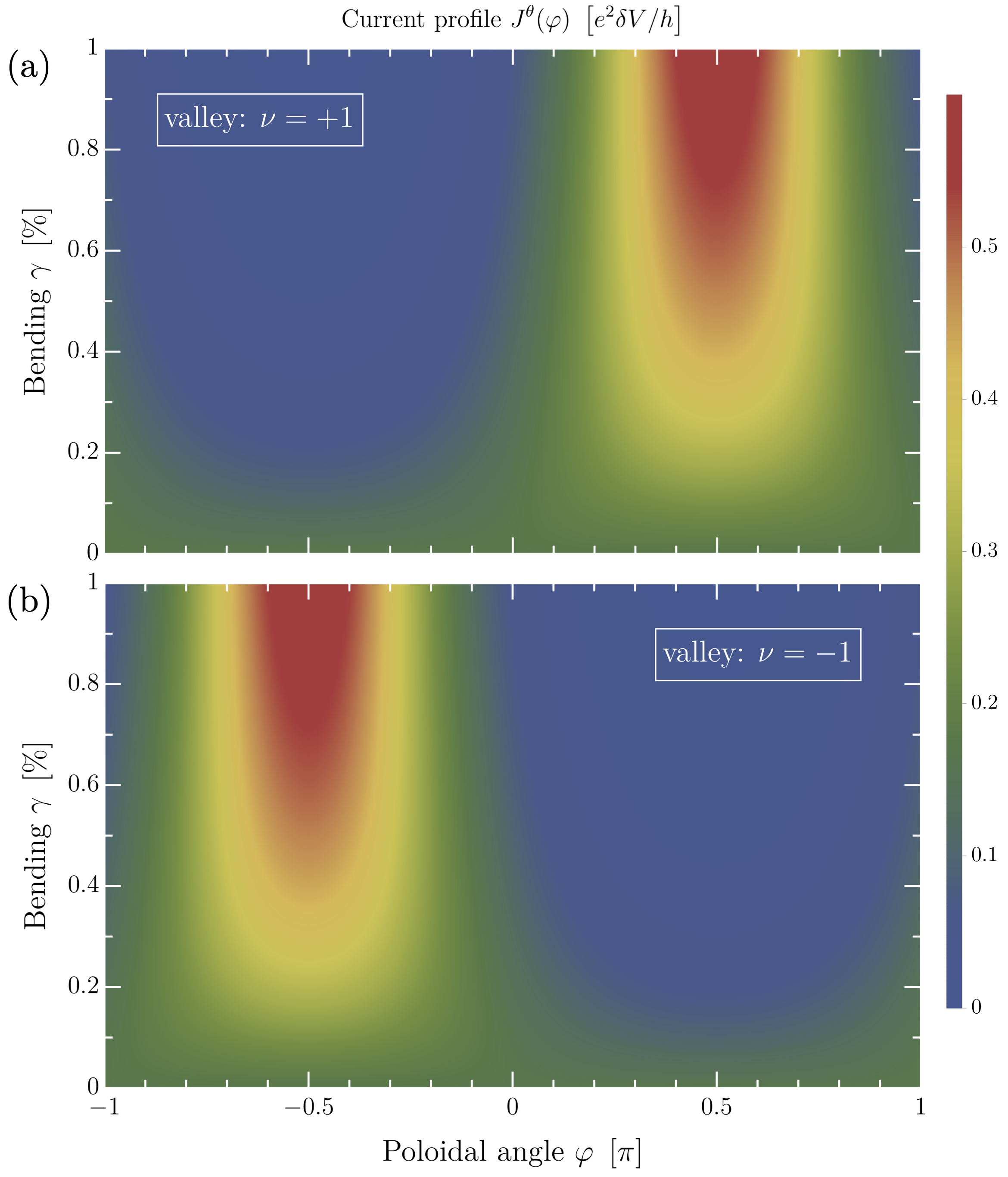}
		\caption{The current profile $J^\theta(\varphi)$ for an armchair nanotube with valley (a) $\nu=+1$ and (b) $\nu=-1$ as a function of the poloidal angle $\varphi$ and the bending parameter $\gamma$. For $\gamma=0$, the current profile is flat. For $\gamma>0$ the current is concentrated on the inner (outer) side of the torus for the valley $\nu=+1$ ($\nu=-1$). The parameters are $\rho\approx47.7\,d_0$ and $\eps_0=0.03\,t_0$ which correspond to the wave numbers $\delta k=1.33\,\rho^{-1}$ and $\delta k=1.38\,\rho^{-1}$ for $\nu=+1$ and $\nu=-1$, respectively. 
		 }\label{fig:4}
\end{figure}

\subsection{Current}
The resulting current density flowing through the bent nanotube is calculated from the solutions $\underline{\psi}$ of the Dirac equation~\eqref{eq:dirac} via
\begin{align}
j^i=\underline{\psi}^\dagger\sigma^a {e}_a^{~i} \underline{\psi} \label{a_strom}
\end{align}
and fulfills the covariant continuity equation
\begin{align}
\nabla_i j^i=\frac{1}{\sqrt{\det \h g }}\left(\partial_i {\sqrt{\det \h g }\, j^i}\right)=0.
\end{align}
In fact, it is the quantity
\begin{align}
J^i=J_0\sqrt{\det \h g }\,\underline{\psi}^\dagger \sigma^a {e}_a^{~i} \underline{\psi}=J_0 \,\underline{\chi}^\dagger \sigma^a {e}_a^{~i} \underline{\chi}, \label{eq:dirac_current}
\end{align}
which should be compared with the actual transport calculations on the discrete lattice~\cite{stegmann_2016}. In the last step, we used the scaling transformation $\underline{\psi}=e^{-S}\underline{\Phi}$. The parameter $J_0$ is chosen such that the electron current has proper dimensions and fulfills $\int_{-\pi}^{\pi} \mathrm{d}\varphi J^\theta(\varphi)=e^2 \delta V/h$, i.e., each Dirac mode $\underline{\psi}$ contributes with the conductance quantum $e^2/h$ times the voltage $\delta V$.
For the zero mode,
we obtain
\begin{align}
J^\theta (\varphi) &\sim  \left[1+\beta\gamma \sin(\varphi)\right]  \ce\left(a_{0},q,\frac{\varphi+\varphi_0}{2}\right) \label{eq:mathieu_current} \\
&\quad \times  \ce\left(a_{0},-q,\frac{\varphi-\varphi_0}{2}\right),  \nonumber\\ 
J^\varphi(\varphi)&=0.
\end{align}

In Fig.~\ref{fig:4}  (a) and (b), the current $J^\theta(\varphi)$ carried by the zero mode for the armchair edge is shown as a function of the poloidal angle $\varphi$ and the bending parameter $\gamma$ for two different valleys $\nu=+1$ and $\nu=-1$, respectively. We find a remarkable feature: While for a straight nanotube, $\gamma=0$, the current for both valleys is homogeneously distributed over the complete surface, we find that as soon as the nanotube is bent, one valley current ($\nu=1$) flows mainly on the inner side while the other valley current ($\nu=-1$) flows mainly on the outer side of the torus. The effect is surprisingly strong considering the bending is not larger than $\gamma=1\,\%$.
A vivid explanation can be given in terms of a pseudo Lorentz force, where according to the left-hand rule electrons are accelerated perpendicular to the velocity and the pseudo-magnetic field. Since the velocity points in toroidal direction and the pseudo-magnetic field $\vb B^{\nu}$ points down (up) for $\nu=+1$ ($\nu=-1$), see Fig.~\ref{fig:1}, the electrons are directed to the inner (outer) side of the torus.  

In the following, these current profiles will be compared with the lattice current profiles obtained from transport calculations using the non-equilibrium Green's function method.

\section{Electron transport}\label{sec:keldysh}
\begin{figure}[h]
	\includegraphics[width=8.6cm]{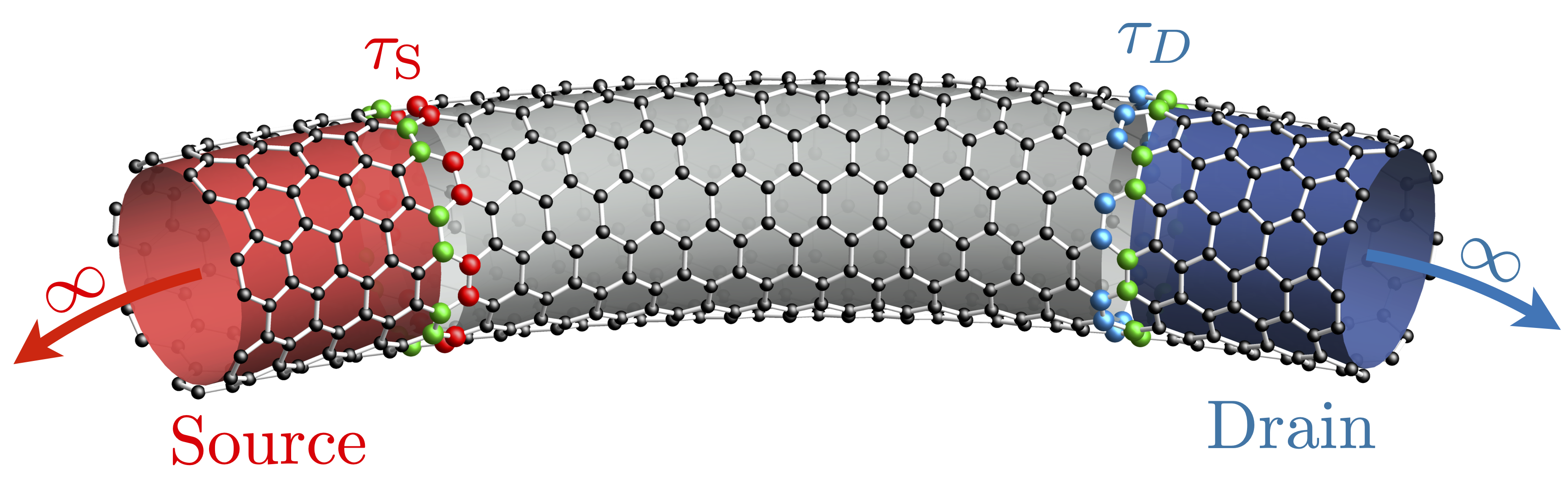}
	\caption{Sketch of a bent nanotube (gray segment) coupled to an electron source (red segment) and drain (blue segment). The electrodes are modeled by semi-infinite nanotubes. The contact points of the system are indicated by red (blue) sites for source (drain). Green sites indicate the surface points of the electrodes.}
	\label{fig:5}
\end{figure}
Finally, we come back to the discrete model, Eq.~\eqref{eq:tightbinding}, to study the electron transport through the bent nanotube (gray) by coupling it to an electron source (red) and drain (blue), see Fig.~\ref{fig:5}.
Here, the leads are modeled by semi-infinite nanotubes. First, we consider $\gamma\neq 0$ where the leads have the same bending as the interior segment. Then, we also briefly discuss straight nanotubes with $\gamma=0$ as leads and we find no qualitative change of the results.

To calculate the local bond current of electrons $J_{\vb n \vb m}$ from site $\vb m$ to site $\vb n$, we use the non-equilibrium Green's function approach in the Keldysh formalism~\cite{caroli_1971,cresti_2003,zienert_2010,lewenkopf_2013,settnes_2015}
\begin{align}\label{eq:curr}
J_{\vb n \vb m}=- {2e}\, \Re \int\frac{\mathrm{d}\omega}{2\pi} \,t_{\vb n\vb m}\,G^<_{\vb n\vb m}(\omega),
\end{align}
with the lesser Green's function $G^<$ which can be calculated using the Keldysh formula $G^<= G^R \Sigma^< (G^R)^\dagger$. Here, the retarded Green's function is given by $G^{R}=\left(\omega -H-\Sigma^R\right)^{-1}$ with $H$ being the Hamiltonian of Eq.~\eqref{eq:tightbinding}.
Now, the problem is reduced to finding the lesser $\Sigma^<$ and the retarded self energy $\Sigma^R$ which effectively simulate the openness of the quantum system through the coupled leads. 
They can be obtained via
\begin{align}
\Sigma^<(\omega)&=\sum_{r=S,D}\tau^\dagger_{r} g_{r}^<(\omega) \tau_{r}, \\
\Sigma^R(\omega)&=\sum_{r=S,D}\tau^\dagger_{r} g_{r}^R(\omega) \tau_{r},
\end{align}
which describe electrons first tunneling with tunneling matrix $\tau_{r}$ from the contact points (red and blue atoms) to the lead $r$ (green atoms), then propagating with the free (without coupling) lead Green's functions $g_r^<$ and $g_r^R$, and then tunneling back into the system with $\tau_r^\dagger$. The contributions of source (S) and drain (D) simply add up. 
The free lead Green's functions are given by
\begin{align}
g_r^< (\omega)&= 2\pi i\, \delta\left(\omega{-}H_r\right)\, n_r, \label{eq:gless}\\
g^R_r(\omega)&=\frac{1}{\omega-H_r+i0^+},
\end{align}
where $H_r$ is the single-particle Hamiltonian of the lead $r$ with eigenvalues $\epsilon_{r,n}(k)$ and eigenstates $\ket{n,k}_r$. To ensure perfect transmission, we assume a perfect translational invariance in the $\theta$-direction, i.e., the leads have the same strain as the system Hamiltonian. 
In this way, we avoid finite-size effects such as standing waves or edge states~\cite{marganska_2011} occuring, e.g., in the wide-band model~\cite{zienert_2010}.
The matrix $n_r$ describes the occupation in the lead $r$ and is defined by $\mel{n,k}{n_r}{n^\prime,k^\prime}_r=\ev{c^\dagger_{r,n^\prime,k^\prime}c_{r,n,k}}_r$, where $c^\dagger_{r,n,k}$ and $c_{r,n,k}$ are the creation and annihilation operators of electrons in the eigenstates $\ket{n,k}_r$ in the bath $r$. 
Thus, $g_r^<$ contains information about the occupations whereas $g_r^R$ contains information about the spectrum of the leads.
Note that in order to determine $\Sigma_<$ and $\Sigma_R$ only the surface Green's functions (at the green sites) are actually needed which for quasi one dimensional systems can be obtained using a recursive scheme, see Appendix~\ref{app:surface_gf} for details~\cite{teichert_2018}.
In the following, we discuss two different transport scenarios.

%%%%%%%%%%%%%%%%%%%%%%%%%%%%%%%%%%%%%%%%%%%%%%%%%%%%
\subsection{Leads each in thermodynamic equilibrium}

We assume that each electrode is in thermodynamic equilibrium such that the occupation matrix $n_r$ is given by the Fermi function
\begin{align}
n_r=f(H_r-\mu_r)=\frac{1} {e^{(H_r-\mu_r)/(k_\text{B} T)}+1},
\end{align}
with the Boltzmann constant $k_\text{B}$, the temperature $T$ and the electrochemical potential $\mu_r$. 
The lesser Green's function $g_r^<$ is then completely determined by the spectrum and thus can be expressed through $g_r^R$ via
\begin{align}
g_r^<(\omega)&=2\pi i\, \delta (\omega-H_r)\, f(\omega-\mu_r),\\
&= -\left[g_r^R(\omega)-(g_r^R)^\dagger(\omega) \right]\, f(\omega-\mu_r).
\end{align}
In the following, we choose $\mu_S=\mu+e\, \delta V$ and $\mu_D=\mu$. For zero temperature ($T \rightarrow 0$), we find the following current 
\begin{align}
J_{\vb n \vb m}&=- {2e}\, \Re \int\limits_\mu^{\mu+e\delta V}\frac{\mathrm{d}\omega}{2\pi}t_{\vb n\vb m}\left[G^R\Sigma^<_S (G^R)^\dagger\right]_{\vb n \vb m}(\omega)\\
&\approx \frac{e^2 \delta V}{\pi}\, \Im \,t_{\vb n\vb m}\left[G^R\Gamma_S (G^R)^\dagger\right]_{\vb n \vb m}(\mu), \label{eq:curr_thermo}
\end{align}
where only energies in the bias window $\mu<\omega<\mu+e\,\delta V$ contribute since the left and the right moving currents with $\omega<\mu$ compensate each other exactly. In the second line, we enter the linear response regime (small bias voltages $\delta V$)  and introduce the coupling matrix
\begin{align}
\Gamma_S=i \left[\Sigma^R_S-(\Sigma^R_S)^\dagger\right]=2\pi\,\tau_S^\dagger \delta(\mu-H_S)\,\tau_S,
\end{align}
which describes the injected electrons. Note that the delta function $\delta(\mu-H_S)$ ensures that only those states with energy $\epsilon_{S,n}(k)=\mu$ are considered.

\begin{figure}[t]
	\includegraphics[width=.49\textwidth]{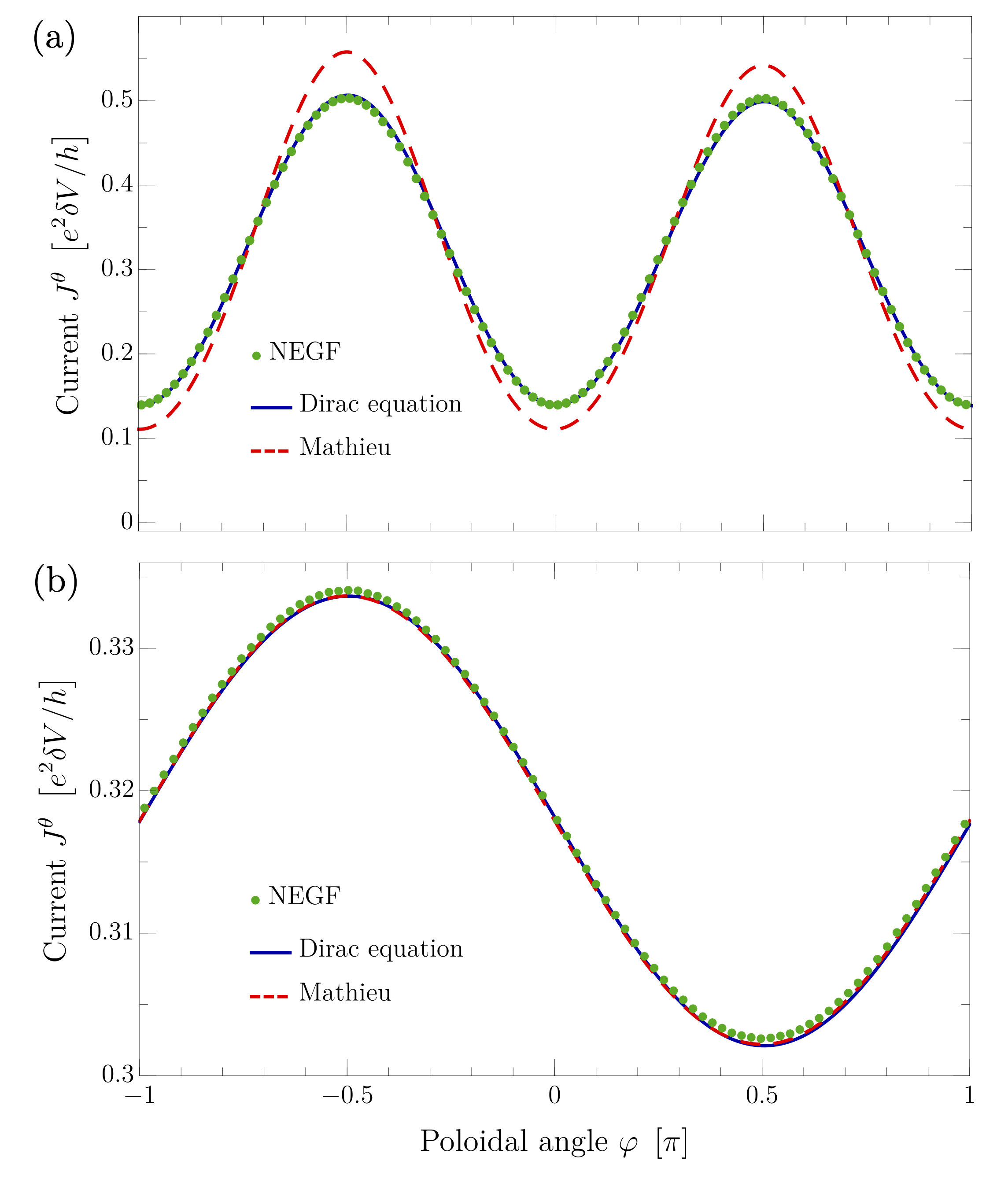}
	\caption{Current profile for an (a) armchair and (b) zigzag nanotube with source and drain each in thermodynamic equilibrium. The numerical NEGF results (green dots) are compared with the numerically solved Dirac equation (blue line) and the analytical solutions in terms of Mathieu functions (red dashed line). The parameters are $\gamma=1\,\%$, $\rho\approx47.7\, d_0$, and (a) $\mu=0.03\,t_0$ and (b) $\mu=0.025\,t_0$. For the NEGF calculations, we use the tube length $L\approx 695\,d_0$ corresponding to approximately $160 000$ atoms.}\label{fig:6} 
\end{figure}
In Fig.~\ref{fig:6}, we employed Eq.~\eqref{eq:curr_thermo} to calculate the current profile $J^\theta(\varphi)$ through (a)  an armchair and (b) a zigzag nanotube at electrochemical potential $\mu$ chosen such that the current is carried only by the two zero modes ($n=0$), one in each valley $\nu=\pm1$.
For the NEGF approach (green dots), we calculated
\begin{align}
J^\theta(\varphi) = \vb{e}_1^{(0)}\cdot \sum_{\vb{n},\vb{m}\in {\cal C} }\frac{\vb n -\vb m}{\vert \vb n -\vb m\vert} J_{\vb n \vb m} ,
\end{align}
i.e. we performed a vector average of the local bond current $J_{\vb n \vb m}$ over the six links $\vb n \leftrightarrow \vb m$ of each hexagonal carbon ring ${\cal C}$ and associate it with the angles $\theta$ and $\varphi$. 
We find that the current is translationally invariant in the $\theta$-direction and thus we only show the $\varphi$ dependence.
In the armchair case, we clearly see that the current is very inhomogeneous and is concentrated both on the inner and the outer side of the torus. 
In the zigzag case, in contrast, the profile is still rather flat but slightly favors the outer side of the torus. 

The results are compared with the current Eq.~\eqref{eq:dirac_current} obtained from the numerically (blue line) and analytically (red dashed line) solved Dirac equation. Since in the transport calculation both valleys $\nu$ are injected at the source electrode we have to consider the incoherent sum of the valley polarized currents
\begin{align}
J^\theta(\varphi)=J^\theta_{\nu=+1}(\varphi) +J^\theta_{\nu=-1}(\varphi).
\end{align}
We find that the transport results are in perfect agreement with the numerically solved Dirac equation~\eqref{eq:dirac_armchair} and \eqref{eq:dirac_zigzag} for the armchair and zigzag case, respectively.
However, while in the zigzag case the analytical solution from Eq.~\eqref{eq:mathieu_current} works equally well, in the armchair case we see small deviations due to the approximation made in
Eq.~\eqref{eq:mathieu_approx}. 

For the armchair case, we can conclude that the pseudo-magnetic field $\vb B^{(\nu)}$ is responsible for a large current splitting on the inner and outer side of the torus.
In contrast, for the zigzag case, we find that the absence of a pseudo-magnetic field leads to a much smaller effect. Nonetheless, the preference of the current (for both valleys) on the outer side is a pure curvature effect and can be explained in terms of the semiclassical trajectories of electrons described by geodesics~\cite{stegmann_2016,Stegmann2018,Ortiz2022}. Using the geodesic deviation equation~\cite{wald2010general}, it immediately becomes apparent that the positive (negative) curvature on the outer (inner) side of the torus is attractive (repelling) for the geodesics.

To confirm the results from Fig.~\ref{fig:4} with certainty, where only the valley current $\nu=+1$ flows on the inner side and only the valley current $\nu=-1$ flows on the outer side of the torus, we next prepare the source in a valley-polarized state.

\subsection{Valley-polarized leads}
\begin{figure}[h]
	\includegraphics[width=8.6cm]{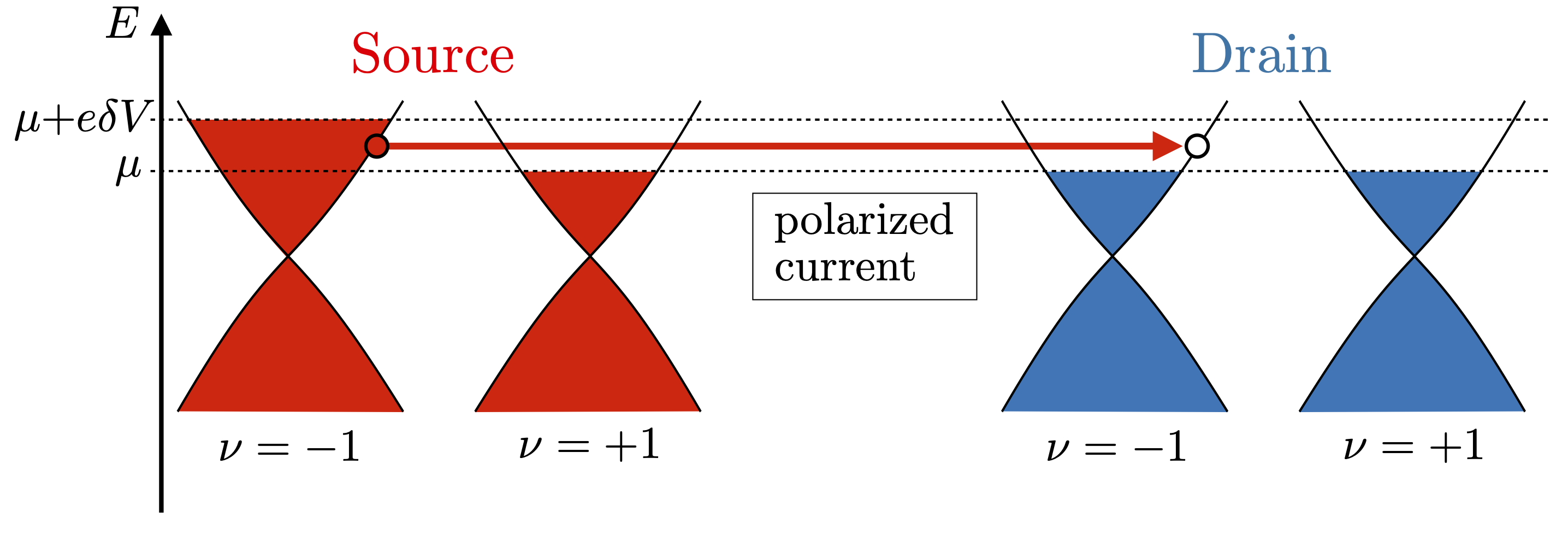}
	\caption{Sketch of the occupation in source (red) and drain (blue). Here, the electrochemical potentials for the source are $\mu_S^{(-1)}=\mu+e\delta V$ and $\mu_S^{(+1)}=\mu$. For the drain we choose $\mu_D^{(-1)}=\mu_D^{(+1)}=\mu$ leading to a valley ($\nu=-1$) polarized current in the bias window.}
	\label{fig:7}
\end{figure}

\begin{figure}[t]
	\includegraphics[width=.49\textwidth]{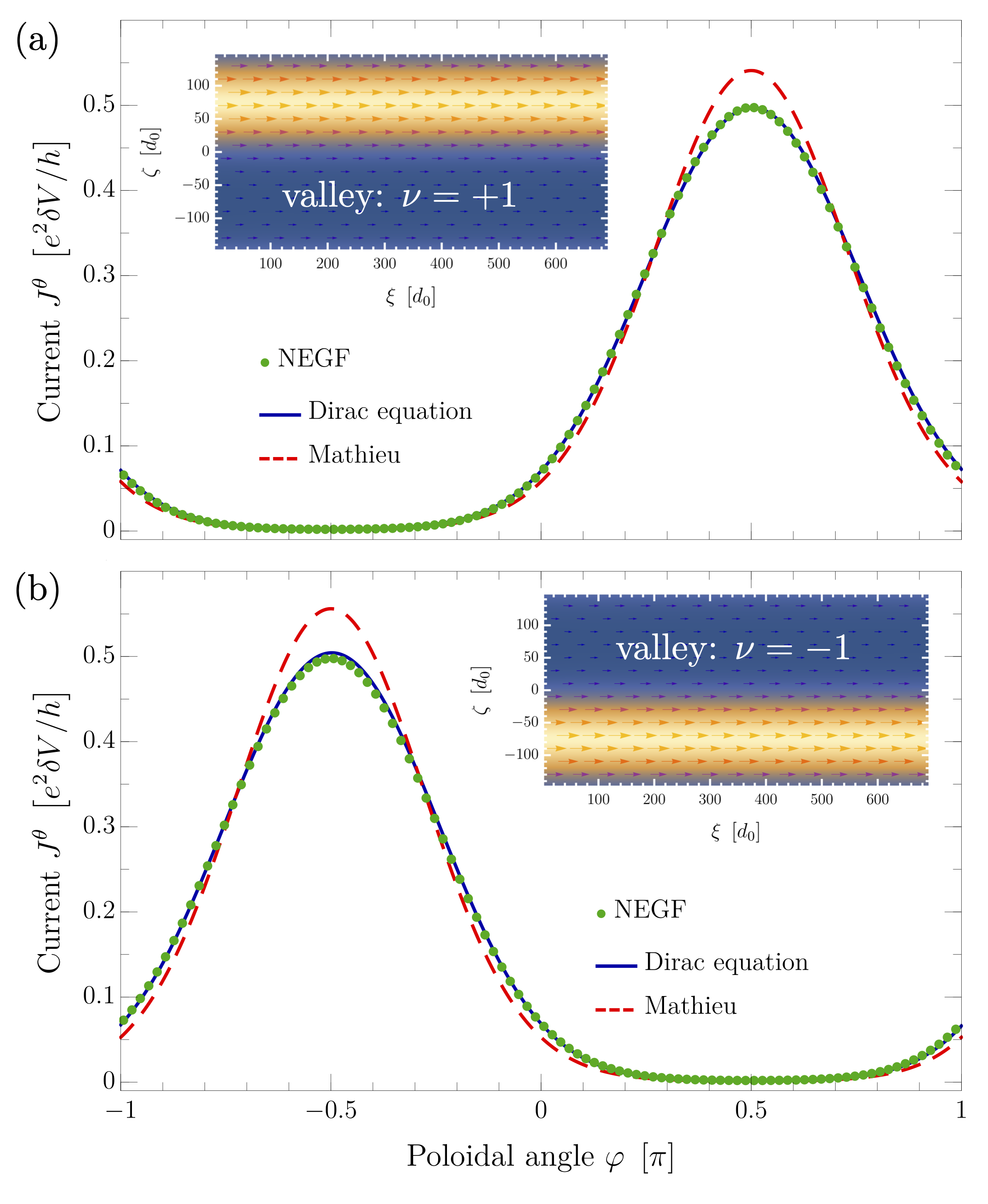}
	\caption{Current profile for an armchair nanotube, where only electrons in the valley (a) $\nu=+1$ and (b) $\nu=-1$ are injected. The numerical NEGF results (green dots) are compared with the numerically solved Dirac equation (blue line) and the approximate analytical solutions in terms of the Mathieu functions (red dashed line). The insets show the NEGF current vector field (red arrows) evaluated on the full nanotube, where the background color indicates its absolute value. The parameters are $\gamma=1\,\%$, $\rho\approx47.7\, d_0$, and $\mu=0.03\,t_0$. For the NEGF calculations, we use a tube length $L\approx 695\,d_0$ corresponding to approximately $160 000$ atoms.
	(For a visualization of the current on the nanotube see Fig.~\ref{fig:1})
	}\label{fig:8}
\end{figure}

\begin{figure}[t]
	\includegraphics[width=.49\textwidth]{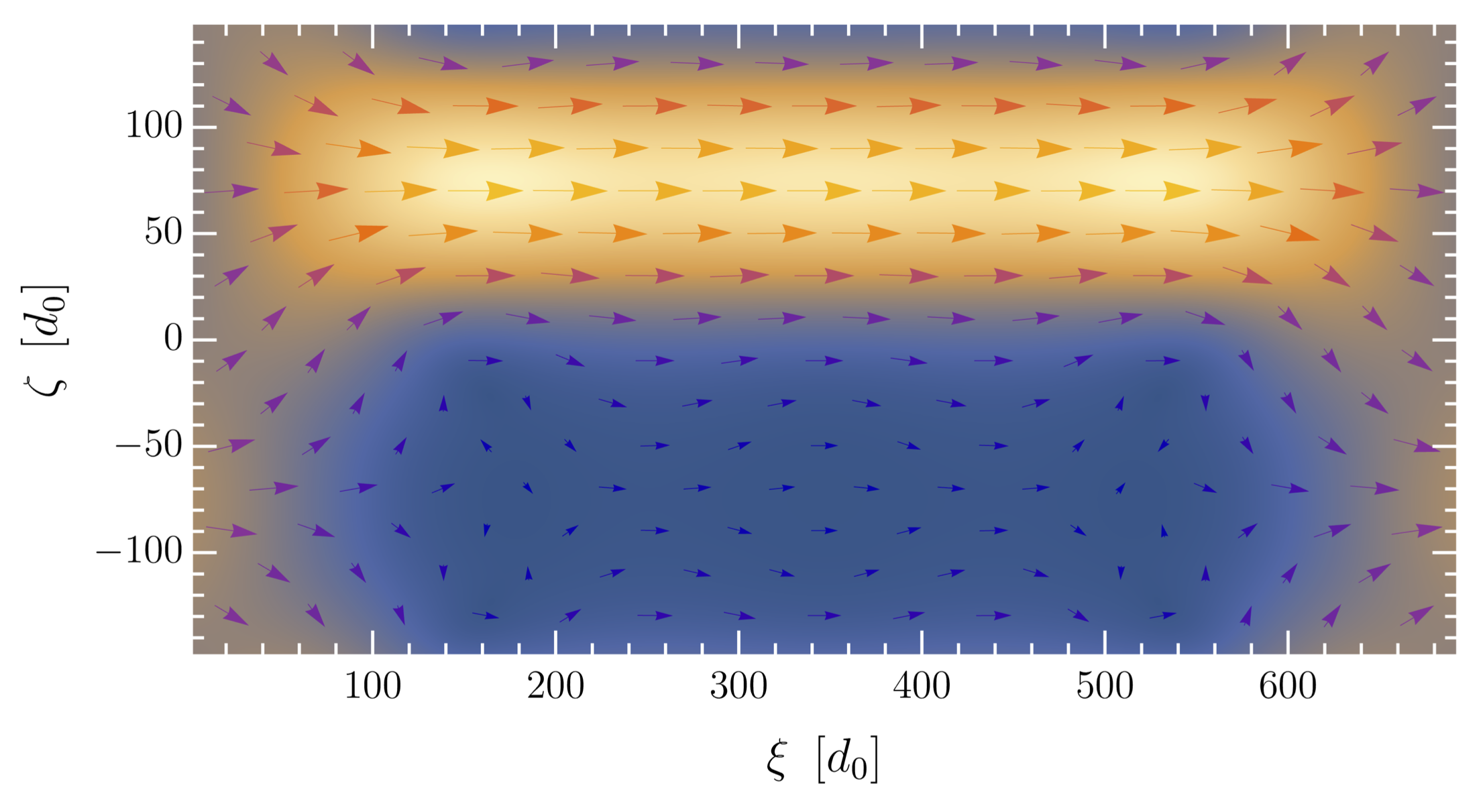}
	\caption{Local current for an armchair nanotube, where source and drain are straight cylinder nanotubes ($\gamma=0\,\%$) and only electrons with valley $\nu=+1$ are injected. The color indicates the absolute value. The parameters are $\gamma=1\,\%$, $\rho\approx47.7\, d_0$, and $\mu=0.028\,t_0$. We use a tube length of $L\approx 695\,d_0$ corresponding to approximately $160 000$ atoms.}\label{fig:9}
\end{figure}

If the leads are in thermal equilibrium, the occupation $n_r$ is valley unpolarized since both degenerate valleys, $\nu=\pm1$, always occur as a total mixture in Eq.~\eqref{eq:curr_thermo}. However, with Eq.~\eqref{eq:gless} we can choose any nonthermal occupation $n_r$ of the leads. Here, we are interested in valley-polarized leads
\begin{align}
n_r&=\sum_{\nu=\pm 1}P^{(\nu)}f\left(H_r-\mu_{r}^{(\nu)}\right)P^{(\nu)},
\end{align}
where $P^{(\nu)}$ is the projector onto the valley $\nu$.
We introduce for each valley a separate electrochemical potential $\mu_r^{(\nu)}$. 
To obtain a valley $\nu$ polarized current, we choose
for the source electrode $\mu_S^{(\nu)}=\mu+e \delta V$ and $\mu_S^{(-\nu)}=\mu$ and for the (unpolarized) drain electrode $\mu_D^{(\nu)}=\mu_D^{(-\nu)}=\mu$, see Fig.~\ref{fig:7}.
Then, we can calculate the current for $T\rightarrow 0$ with Eq.~\eqref{eq:curr_thermo} by using the simple replacement 
\begin{align}\label{eq:curr_polarized}
\Gamma_S~ \rightarrow ~\Gamma_S^{(\nu)}=i P^{(\nu)}\left[\Sigma^R_S-(\Sigma^R_S)^\dagger\right]P^{(\nu)}.
\end{align}

Now we have the tool to precisely inject only one valley $\nu$ into the nanotube. The results are shown in Fig.~\ref{fig:8}. The insets show the current vector field (averaged over carbon rings) on the full nanotube, where the color indicates its absolute value. Again, we find for the current profiles a perfect agreement between the NEGF results (green dots) and the Dirac equation (blue line), while there are small deviations from the analytical solution (red dashed line).  
Moreover, we can confirm that the bent nanotube works as a valley splitter.
If electrons in the valley $\nu=+1$ ($\nu = -1$) are injected the current flows only on the inner (outer) side of the bent nanotube. 
In fact, we find an inner-to-outer current contrast of
\begin{align}
\frac{\max J^\theta_\nu - \min J^\theta_\nu}{\max J^\theta_\nu}\approx 99.6\,\%
\end{align}
for a bending parameter of only $\gamma=1\,\%$.

Finally, in order to better understand the influence of the form of the electrodes on the local current flow, in Fig.~\ref{fig:9}, we repeat the calculations with source and drain now being straight nanotubes ($\gamma=0\,\%$). 
Moreover, we modify the system such that the bending is smoothly increased from $\gamma=0\,\%$ at the  contacts to $\gamma=1\,\%$ in the interior of the nanotube with a characteristic length scale of $100\,d_0$.
We inject electrons in the zero mode of one valley, $\nu=+1$, which for the cylinder nanotube corresponds to a flat current profile, see Fig.~\ref{fig:4}. 
Nonetheless, we find that as the current reaches the region with the bending of $\gamma=1\,\%$ it again becomes fully localized on the inner side of the torus, similarly as in the case with bent electrodes. Therefore, we conclude that the form of the electrodes has no significant impact on the current flow profile in the interior of the bent nanotube.

\section{\label{sec:conclusions} Conclusions}
In this paper, we describe the low-energy excitations of a torus shaped nanotube by an effective Dirac equation in curved space that includes a coupling to a strain-induced pseudo-magnetic field. We find that the approximate solutions are given by the Mathieu functions. In particular, the nanotube with an armchair edge induces a pseudo-magnetic field perpendicular to the plane of the torus and thus acts as a strong valley splitter, where the current carried by the zero mode favors electrons of one valley on the inner side and of the other valley on the outer side of the torus. The effect is surprisingly strong leading to an inner-to-outer current contrast of $99.6\,\%$ for a small strain of $\gamma=1\,\%$. In contrast, the zigzag nanotubes are largely unaffected by the bending due to the absence of the pseudo-magnetic field. We compare the analytical solutions with the results of transport calculations utilizing the non-equilibrium Green's function method in the Keldysh formalism. In order to inject the electrons only in one valley from the source, we prepare the lead in a valley-polarized state. 
In all cases, we get a strong quantitative agreement and thus we confirm the validity of the effective continuous description of bent carbon nanotubes based on the Dirac equation in curved space.

Our findings will be important for electronic devices based on carbon nanotubes and may lead  to new applications in nanoelectronics such as valley splitters. In our future work, we plan to address the question of how different chiralities, either embedded in the atomic structure (beyond the AC and ZZ cases) or in form of an elastic twist of the nanotube, affect the local current flow.

\begin{acknowledgments}
We thank Dr.~Magdalena Margańska for helpful discussions. 
This work was supported by the Deutsche Forschungsgemeinschaft (DFG, German  Research Foundation) under Project-ID 278162697 -- SFB 1242, UNAM-PAPIIT under Project-ID IN103922 and CONACYT under Project-ID A1-S-13469.
\end{acknowledgments}

%%%%%%%%%%%%%%%%%%%%%%%%%%%%%%%%%%%%%%%%%%%%%%%%%%%%%%%%%%%%%%
\appendix
\section{Spin connection}\label{app:spin_connection}
The spin connection $\Omega_j$ that enters the Dirac equation~\eqref{eq:dirac} describes how the zweibein $e_a^{~i}(\vb{x})$ has to be transformed from one point to another to maintain an orthonormal frame. It is defined as
\begin{align}
{\Omega_j}=\frac{1}{8} w_{jab}\lbrack\sigma^a,\sigma^b\rbrack,
\end{align}
with the rotation coefficients
\begin{align}
w_{jab}={e}_{al}\nabla_j {e}_b^{~l}=g_{ml}{e}_a^{~m}(\partial_j {e}_b^{~l}+{\Gamma}_{jk}^l {e}_b^{~k}).
\end{align}
Here, ${\Gamma}_{jk}^l $ denote the Christoffel--Symbols ${\Gamma}_{ij}^k=\frac{1}{2}{g}^{kl}\left( \partial_i {g}_{lj}+\partial_j{g}_{il}-\partial_l {g}_{ij} \right)$ 
of which only
\begin{align}
  {\Gamma}_{\xi\xi}^{\zeta}&=\frac{\beta \gamma}{\rho}\Cose, \\
  {\Gamma}_{\xi\zeta}^{\xi}&={\Gamma}_{\zeta\xi}^{\xi}=-\frac{\beta \gamma}{\rho}\Cose,
\end{align}
are non-zero. Putting all together, we find for the nonzero elements
\begin{align}
w_{\xi21}=-w_{\xi12}=\frac{\beta{\gamma}}{\rho}\Cose,
\end{align}
and thus for the spin connection
\begin{align}
\Omega_\xi=\mp \frac{i\nu}{2} w_{\xi 21} \,\sigma_z \quad \text{and} \quad  \Omega_\zeta&=0,
\end{align}
with the Pauli matrix $\sigma_z$. The upper and lower sign is used for the armchair and zigzag edges, respectively. The corresponding term in the Dirac equation~\eqref{eq:dirac} becomes 
\begin{align}
\sigma^a e^{~j}_a \Omega_j&=\mp \frac{i\nu}{2} w_{\xi 21} e^{~\xi}_1\sigma^1\sigma_z\\
&= -\frac{1}{2} w_{\xi 21} e^{~\xi}_1\sigma^2\\
&\approx \sigma^2e^{~\zeta}_2\left(-\frac{1}{2} w_{\xi 21} \right) =\sigma^a e^{~j }_a {b_j},
\end{align}
where we approximated the expression in leading order in $\gamma$ by using $e^{~\xi}_1\approx e^{~\zeta}_2=1$. Using the product $\sigma^1 \sigma_z=\mp i \nu \sigma^2$, we find that the spin connection effectively acts like a real vector field
\begin{align}
{b_\xi}=0 \quad \text{and} \quad  b_{\zeta}=-\frac{\beta{\gamma}}{2\rho}\Cose,
\end{align}
which can be written as a gradient of a scalar field $b_j=\partial_j S$ with $S=\log \left( \sqrt[4]{\det \h g}\right)$.
Finally, the field $b_j$ can be completely removed from the Dirac equation~\eqref{eq:dirac} by the scaling transformation $\underline{\Phi}=e^S\underline{\psi}$.

\section{Surface Green's function}\label{app:surface_gf}

Since the system is quasi one dimensional, we can use the Hamiltonian $h_R$ of a unit ring (cf. the green atoms in the insets of Fig.~\ref{fig:2})  that couples with $\tau$ ($\tau^\dagger$) to the neighboring unit ring to the right (left).
Then we can employ the recursive relations
\begin{align}
g^R_{S,n}=(\omega - h_R + \tau g^R_{S,n-1} \tau^\dagger)^{-1}, \\
g^R_{D,n}=(\omega - h_R + \tau^\dagger g^D_{S,n-1} \tau)^{-1},
\end{align}
which relate the retarded surface Green's function ( $g^R_{r,n}$) of lead $r$ comprised from $n$ unit rings to the one ($g^R_{r,n-1}$) comprised from $n-1$ unit rings. 
For one ring $n=1$, we get $g^R_{S,1}=g^R_{D,1}=(\omega - h_R +i 0^+)^{-1}$.  
Each time the recursion relation is applied, a unit ring is added to source and drain. In practice, to simulate semi-infinite leads ($n\rightarrow \infty$),
the iteration is stopped when the surface Green's function is sufficiently converged.

%%%%%%%%%%%%%%%%%%%%%%%%%%%%%%%%%%%%%%%%%%%%%%%%%%%%%%%%%%%%%%%%%%%%%%%%%%%%%%%%%%%%%%%%%%%%%%%%%%%%
%%%%%%%%%%%%%%%%%%%%%%%%%%%%%%%%%%%%%%%%%%%%%%%%%%%%%%%%%%%%%%%%%%%%%%%%%%%%%%%%%%%%%%%%%%%%%%%%%%%%
\bibliography{References}

\begin{thebibliography}{90}%
\makeatletter
\providecommand \@ifxundefined [1]{%
 \@ifx{#1\undefined}
}%
\providecommand \@ifnum [1]{%
 \ifnum #1\expandafter \@firstoftwo
 \else \expandafter \@secondoftwo
 \fi
}%
\providecommand \@ifx [1]{%
 \ifx #1\expandafter \@firstoftwo
 \else \expandafter \@secondoftwo
 \fi
}%
\providecommand \natexlab [1]{#1}%
\providecommand \enquote  [1]{``#1''}%
\providecommand \bibnamefont  [1]{#1}%
\providecommand \bibfnamefont [1]{#1}%
\providecommand \citenamefont [1]{#1}%
\providecommand \href@noop [0]{\@secondoftwo}%
\providecommand \href [0]{\begingroup \@sanitize@url \@href}%
\providecommand \@href[1]{\@@startlink{#1}\@@href}%
\providecommand \@@href[1]{\endgroup#1\@@endlink}%
\providecommand \@sanitize@url [0]{\catcode `\\12\catcode `\$12\catcode
  `\&12\catcode `\#12\catcode `\^12\catcode `\_12\catcode `\%12\relax}%
\providecommand \@@startlink[1]{}%
\providecommand \@@endlink[0]{}%
\providecommand \url  [0]{\begingroup\@sanitize@url \@url }%
\providecommand \@url [1]{\endgroup\@href {#1}{\urlprefix }}%
\providecommand \urlprefix  [0]{URL }%
\providecommand \Eprint [0]{\href }%
\providecommand \doibase [0]{https://doi.org/}%
\providecommand \selectlanguage [0]{\@gobble}%
\providecommand \bibinfo  [0]{\@secondoftwo}%
\providecommand \bibfield  [0]{\@secondoftwo}%
\providecommand \translation [1]{[#1]}%
\providecommand \BibitemOpen [0]{}%
\providecommand \bibitemStop [0]{}%
\providecommand \bibitemNoStop [0]{.\EOS\space}%
\providecommand \EOS [0]{\spacefactor3000\relax}%
\providecommand \BibitemShut  [1]{\csname bibitem#1\endcsname}%
\let\auto@bib@innerbib\@empty
%</preamble>
\bibitem [{\citenamefont {Novoselov}\ \emph {et~al.}(2005)\citenamefont
  {Novoselov}, \citenamefont {Geim}, \citenamefont {Morozov}, \citenamefont
  {Jiang}, \citenamefont {Katsnelson}, \citenamefont {Grigorieva},
  \citenamefont {Dubonos},\ and\ \citenamefont {Firsov}}]{Novoselov+Geim2005}%
  \BibitemOpen
  \bibfield  {author} {\bibinfo {author} {\bibfnamefont {K.~S.}\ \bibnamefont
  {Novoselov}}, \bibinfo {author} {\bibfnamefont {A.~K.}\ \bibnamefont {Geim}},
  \bibinfo {author} {\bibfnamefont {S.~V.}\ \bibnamefont {Morozov}}, \bibinfo
  {author} {\bibfnamefont {D.}~\bibnamefont {Jiang}}, \bibinfo {author}
  {\bibfnamefont {M.~I.}\ \bibnamefont {Katsnelson}}, \bibinfo {author}
  {\bibfnamefont {I.~V.}\ \bibnamefont {Grigorieva}}, \bibinfo {author}
  {\bibfnamefont {S.~V.}\ \bibnamefont {Dubonos}},\ and\ \bibinfo {author}
  {\bibfnamefont {A.~A.}\ \bibnamefont {Firsov}},\ }\bibfield  {title}
  {\bibinfo {title} {Two-dimensional gas of massless {D}irac fermions in
  graphene},\ }\href {https://doi.org/10.1038/nature04233} {\bibfield
  {journal} {\bibinfo  {journal} {Nature}\ }\textbf {\bibinfo {volume} {438}},\
  \bibinfo {pages} {197} (\bibinfo {year} {2005})}\BibitemShut {NoStop}%
\bibitem [{\citenamefont {Katsnelson}\ and\ \citenamefont
  {Novoselov}(2007)}]{Katsnelson2007}%
  \BibitemOpen
  \bibfield  {author} {\bibinfo {author} {\bibfnamefont {M.}~\bibnamefont
  {Katsnelson}}\ and\ \bibinfo {author} {\bibfnamefont {K.}~\bibnamefont
  {Novoselov}},\ }\bibfield  {title} {\bibinfo {title} {Graphene: New bridge
  between condensed matter physics and quantum electrodynamics},\ }\href
  {https://doi.org/https://doi.org/10.1016/j.ssc.2007.02.043} {\bibfield
  {journal} {\bibinfo  {journal} {Solid State Commun.}\ }\textbf {\bibinfo
  {volume} {143}},\ \bibinfo {pages} {3} (\bibinfo {year} {2007})}\BibitemShut
  {NoStop}%
\bibitem [{\citenamefont {Fialkovsky}\ and\ \citenamefont
  {Vassilevich}(2012)}]{Fialkovsky2012}%
  \BibitemOpen
  \bibfield  {author} {\bibinfo {author} {\bibfnamefont {I.~V.}\ \bibnamefont
  {Fialkovsky}}\ and\ \bibinfo {author} {\bibfnamefont {D.~V.}\ \bibnamefont
  {Vassilevich}},\ }\bibfield  {title} {\bibinfo {title} {Quantum field theory
  in graphene},\ }\href {https://doi.org/10.1142/S0217751X1260007X} {\bibfield
  {journal} {\bibinfo  {journal} {Int. J. Mod. Phys. A}\ }\textbf {\bibinfo
  {volume} {27}},\ \bibinfo {pages} {1260007} (\bibinfo {year}
  {2012})}\BibitemShut {NoStop}%
\bibitem [{\citenamefont {Castro~Neto}\ \emph {et~al.}(2009)\citenamefont
  {Castro~Neto}, \citenamefont {Guinea}, \citenamefont {Peres}, \citenamefont
  {Novoselov},\ and\ \citenamefont {Geim}}]{CastroNeto2009review}%
  \BibitemOpen
  \bibfield  {author} {\bibinfo {author} {\bibfnamefont {A.~H.}\ \bibnamefont
  {Castro~Neto}}, \bibinfo {author} {\bibfnamefont {F.}~\bibnamefont {Guinea}},
  \bibinfo {author} {\bibfnamefont {N.~M.~R.}\ \bibnamefont {Peres}}, \bibinfo
  {author} {\bibfnamefont {K.~S.}\ \bibnamefont {Novoselov}},\ and\ \bibinfo
  {author} {\bibfnamefont {A.~K.}\ \bibnamefont {Geim}},\ }\bibfield  {title}
  {\bibinfo {title} {The electronic properties of graphene},\ }\href
  {https://doi.org/10.1103/RevModPhys.81.109} {\bibfield  {journal} {\bibinfo
  {journal} {Rev. Mod. Phys.}\ }\textbf {\bibinfo {volume} {81}},\ \bibinfo
  {pages} {109} (\bibinfo {year} {2009})}\BibitemShut {NoStop}%
\bibitem [{\citenamefont {Katsnelson}(2020)}]{Katsnelson2020}%
  \BibitemOpen
  \bibfield  {author} {\bibinfo {author} {\bibfnamefont {M.~I.}\ \bibnamefont
  {Katsnelson}},\ }\href {https://doi.org/10.1017/9781108617567} {\emph
  {\bibinfo {title} {The Physics of Graphene}}},\ \bibinfo {edition} {2nd}\
  ed.\ (\bibinfo  {publisher} {Cambridge University Press, Cambridge},\
  \bibinfo {year} {2020})\BibitemShut {NoStop}%
\bibitem [{\citenamefont {Foa~Torres}\ \emph {et~al.}(2020)\citenamefont
  {Foa~Torres}, \citenamefont {Roche},\ and\ \citenamefont
  {Charlier}}]{Torres2020}%
  \BibitemOpen
  \bibfield  {author} {\bibinfo {author} {\bibfnamefont {L.~E.~F.}\
  \bibnamefont {Foa~Torres}}, \bibinfo {author} {\bibfnamefont
  {S.}~\bibnamefont {Roche}},\ and\ \bibinfo {author} {\bibfnamefont {J.-C.}\
  \bibnamefont {Charlier}},\ }\href {https://doi.org/10.1017/9781108664462}
  {\emph {\bibinfo {title} {Introduction to Graphene-Based Nanomaterials: From
  Electronic Structure to Quantum Transport}}},\ \bibinfo {edition} {2nd}\ ed.\
  (\bibinfo  {publisher} {Cambridge University Press, Cambridge},\ \bibinfo
  {year} {2020})\BibitemShut {NoStop}%
\bibitem [{\citenamefont {Gentile}\ \emph {et~al.}(2022)\citenamefont
  {Gentile}, \citenamefont {Cuoco}, \citenamefont {Volkov}, \citenamefont
  {Ying}, \citenamefont {Vera-Marun}, \citenamefont {Makarov},\ and\
  \citenamefont {Ortix}}]{gentile_2022}%
  \BibitemOpen
  \bibfield  {author} {\bibinfo {author} {\bibfnamefont {P.}~\bibnamefont
  {Gentile}}, \bibinfo {author} {\bibfnamefont {M.}~\bibnamefont {Cuoco}},
  \bibinfo {author} {\bibfnamefont {O.~M.}\ \bibnamefont {Volkov}}, \bibinfo
  {author} {\bibfnamefont {Z.-J.}\ \bibnamefont {Ying}}, \bibinfo {author}
  {\bibfnamefont {I.~J.}\ \bibnamefont {Vera-Marun}}, \bibinfo {author}
  {\bibfnamefont {D.}~\bibnamefont {Makarov}},\ and\ \bibinfo {author}
  {\bibfnamefont {C.}~\bibnamefont {Ortix}},\ }\bibfield  {title} {\bibinfo
  {title} {Electronic materials with nanoscale curved geometries},\ }\href
  {https://doi.org/10.1038/s41928-022-00820-z} {\bibfield  {journal} {\bibinfo
  {journal} {Nat. Electron.}\ }\textbf {\bibinfo {volume} {5}},\ \bibinfo
  {pages} {551} (\bibinfo {year} {2022})}\BibitemShut {NoStop}%
\bibitem [{\citenamefont {Ortiz}\ \emph {et~al.}(2022)\citenamefont {Ortiz},
  \citenamefont {Szpak},\ and\ \citenamefont {Stegmann}}]{Ortiz2022}%
  \BibitemOpen
  \bibfield  {author} {\bibinfo {author} {\bibfnamefont {W.}~\bibnamefont
  {Ortiz}}, \bibinfo {author} {\bibfnamefont {N.}~\bibnamefont {Szpak}},\ and\
  \bibinfo {author} {\bibfnamefont {T.}~\bibnamefont {Stegmann}},\ }\bibfield
  {title} {\bibinfo {title} {Graphene nanoelectromechanical systems as
  valleytronic devices},\ }\href {https://doi.org/10.1103/PhysRevB.106.035416}
  {\bibfield  {journal} {\bibinfo  {journal} {Phys. Rev. B}\ }\textbf {\bibinfo
  {volume} {106}},\ \bibinfo {pages} {035416} (\bibinfo {year}
  {2022})}\BibitemShut {NoStop}%
\bibitem [{\citenamefont {Ma\~nes}(2007)}]{manes_2007}%
  \BibitemOpen
  \bibfield  {author} {\bibinfo {author} {\bibfnamefont {J.~L.}\ \bibnamefont
  {Ma\~nes}},\ }\bibfield  {title} {\bibinfo {title} {Symmetry-based approach
  to electron-phonon interactions in graphene},\ }\href
  {https://doi.org/10.1103/PhysRevB.76.045430} {\bibfield  {journal} {\bibinfo
  {journal} {Phys. Rev. B}\ }\textbf {\bibinfo {volume} {76}},\ \bibinfo
  {pages} {045430} (\bibinfo {year} {2007})}\BibitemShut {NoStop}%
\bibitem [{\citenamefont {von Oppen}\ \emph {et~al.}(2009)\citenamefont {von
  Oppen}, \citenamefont {Guinea},\ and\ \citenamefont
  {Mariani}}]{vonoppen_2009}%
  \BibitemOpen
  \bibfield  {author} {\bibinfo {author} {\bibfnamefont {F.}~\bibnamefont {von
  Oppen}}, \bibinfo {author} {\bibfnamefont {F.}~\bibnamefont {Guinea}},\ and\
  \bibinfo {author} {\bibfnamefont {E.}~\bibnamefont {Mariani}},\ }\bibfield
  {title} {\bibinfo {title} {Synthetic electric fields and phonon damping in
  carbon nanotubes and graphene},\ }\href
  {https://doi.org/10.1103/PhysRevB.80.075420} {\bibfield  {journal} {\bibinfo
  {journal} {Phys. Rev. B}\ }\textbf {\bibinfo {volume} {80}},\ \bibinfo
  {pages} {075420} (\bibinfo {year} {2009})}\BibitemShut {NoStop}%
\bibitem [{\citenamefont {Vozmediano}\ \emph {et~al.}(2010)\citenamefont
  {Vozmediano}, \citenamefont {Katsnelson},\ and\ \citenamefont
  {Guinea}}]{vozmediano_2010}%
  \BibitemOpen
  \bibfield  {author} {\bibinfo {author} {\bibfnamefont {M.}~\bibnamefont
  {Vozmediano}}, \bibinfo {author} {\bibfnamefont {M.}~\bibnamefont
  {Katsnelson}},\ and\ \bibinfo {author} {\bibfnamefont {F.}~\bibnamefont
  {Guinea}},\ }\bibfield  {title} {\bibinfo {title} {Gauge fields in
  graphene},\ }\href
  {https://doi.org/https://doi.org/10.1016/j.physrep.2010.07.003} {\bibfield
  {journal} {\bibinfo  {journal} {Phys. Rep.}\ }\textbf {\bibinfo {volume}
  {496}},\ \bibinfo {pages} {109} (\bibinfo {year} {2010})}\BibitemShut
  {NoStop}%
\bibitem [{\citenamefont {Wakker}\ \emph {et~al.}(2011)\citenamefont {Wakker},
  \citenamefont {Tiwari},\ and\ \citenamefont {Blaauboer}}]{wakker_2011}%
  \BibitemOpen
  \bibfield  {author} {\bibinfo {author} {\bibfnamefont {G.~M.~M.}\
  \bibnamefont {Wakker}}, \bibinfo {author} {\bibfnamefont {R.~P.}\
  \bibnamefont {Tiwari}},\ and\ \bibinfo {author} {\bibfnamefont
  {M.}~\bibnamefont {Blaauboer}},\ }\bibfield  {title} {\bibinfo {title}
  {Localization and circulating currents in curved graphene devices},\ }\href
  {https://doi.org/10.1103/PhysRevB.84.195427} {\bibfield  {journal} {\bibinfo
  {journal} {Phys. Rev. B}\ }\textbf {\bibinfo {volume} {84}},\ \bibinfo
  {pages} {195427} (\bibinfo {year} {2011})}\BibitemShut {NoStop}%
\bibitem [{\citenamefont {Mucha-Kruczy{\'n}ski}\ and\ \citenamefont
  {Fal'ko}(2012)}]{mucha_2012}%
  \BibitemOpen
  \bibfield  {author} {\bibinfo {author} {\bibfnamefont {M.}~\bibnamefont
  {Mucha-Kruczy{\'n}ski}}\ and\ \bibinfo {author} {\bibfnamefont
  {V.}~\bibnamefont {Fal'ko}},\ }\bibfield  {title} {\bibinfo {title}
  {Pseudo-magnetic field distribution and pseudo-{L}andau levels in suspended
  graphene flakes},\ }\href
  {https://doi.org/https://doi.org/10.1016/j.ssc.2012.04.035} {\bibfield
  {journal} {\bibinfo  {journal} {Solid State Commun.}\ }\textbf {\bibinfo
  {volume} {152}},\ \bibinfo {pages} {1442} (\bibinfo {year} {2012})},\
  \bibinfo {note} {exploring Graphene, Recent Research Advances}\BibitemShut
  {NoStop}%
\bibitem [{\citenamefont {Kitt}\ \emph {et~al.}(2012)\citenamefont {Kitt},
  \citenamefont {Pereira}, \citenamefont {Swan},\ and\ \citenamefont
  {Goldberg}}]{kitt_2012}%
  \BibitemOpen
  \bibfield  {author} {\bibinfo {author} {\bibfnamefont {A.~L.}\ \bibnamefont
  {Kitt}}, \bibinfo {author} {\bibfnamefont {V.~M.}\ \bibnamefont {Pereira}},
  \bibinfo {author} {\bibfnamefont {A.~K.}\ \bibnamefont {Swan}},\ and\
  \bibinfo {author} {\bibfnamefont {B.~B.}\ \bibnamefont {Goldberg}},\
  }\bibfield  {title} {\bibinfo {title} {Lattice-corrected strain-induced
  vector potentials in graphene},\ }\href
  {https://doi.org/10.1103/PhysRevB.85.115432} {\bibfield  {journal} {\bibinfo
  {journal} {Phys. Rev. B}\ }\textbf {\bibinfo {volume} {85}},\ \bibinfo
  {pages} {115432} (\bibinfo {year} {2012})}\BibitemShut {NoStop}%
\bibitem [{\citenamefont {Neek-Amal}\ \emph {et~al.}(2012)\citenamefont
  {Neek-Amal}, \citenamefont {Covaci},\ and\ \citenamefont
  {Peeters}}]{neek-amal_2012}%
  \BibitemOpen
  \bibfield  {author} {\bibinfo {author} {\bibfnamefont {M.}~\bibnamefont
  {Neek-Amal}}, \bibinfo {author} {\bibfnamefont {L.}~\bibnamefont {Covaci}},\
  and\ \bibinfo {author} {\bibfnamefont {F.~M.}\ \bibnamefont {Peeters}},\
  }\bibfield  {title} {\bibinfo {title} {Nanoengineered nonuniform strain in
  graphene using nanopillars},\ }\href
  {https://doi.org/10.1103/PhysRevB.86.041405} {\bibfield  {journal} {\bibinfo
  {journal} {Phys. Rev. B}\ }\textbf {\bibinfo {volume} {86}},\ \bibinfo
  {pages} {041405(R)} (\bibinfo {year} {2012})}\BibitemShut {NoStop}%
\bibitem [{\citenamefont {Carrillo-Bastos}\ \emph {et~al.}(2014)\citenamefont
  {Carrillo-Bastos}, \citenamefont {Faria}, \citenamefont {Latg\'e},
  \citenamefont {Mireles},\ and\ \citenamefont {Sandler}}]{CarrilloBastos2014}%
  \BibitemOpen
  \bibfield  {author} {\bibinfo {author} {\bibfnamefont {R.}~\bibnamefont
  {Carrillo-Bastos}}, \bibinfo {author} {\bibfnamefont {D.}~\bibnamefont
  {Faria}}, \bibinfo {author} {\bibfnamefont {A.}~\bibnamefont {Latg\'e}},
  \bibinfo {author} {\bibfnamefont {F.}~\bibnamefont {Mireles}},\ and\ \bibinfo
  {author} {\bibfnamefont {N.}~\bibnamefont {Sandler}},\ }\bibfield  {title}
  {\bibinfo {title} {Gaussian deformations in graphene ribbons: Flowers and
  confinement},\ }\href {https://doi.org/10.1103/PhysRevB.90.041411} {\bibfield
   {journal} {\bibinfo  {journal} {Phys. Rev. B}\ }\textbf {\bibinfo {volume}
  {90}},\ \bibinfo {pages} {041411(R)} (\bibinfo {year} {2014})}\BibitemShut
  {NoStop}%
\bibitem [{\citenamefont {Naumis}\ \emph {et~al.}(2017)\citenamefont {Naumis},
  \citenamefont {Barraza-Lopez}, \citenamefont {Oliva-Leyva},\ and\
  \citenamefont {Terrones}}]{Naumis2017}%
  \BibitemOpen
  \bibfield  {author} {\bibinfo {author} {\bibfnamefont {G.~G.}\ \bibnamefont
  {Naumis}}, \bibinfo {author} {\bibfnamefont {S.}~\bibnamefont
  {Barraza-Lopez}}, \bibinfo {author} {\bibfnamefont {M.}~\bibnamefont
  {Oliva-Leyva}},\ and\ \bibinfo {author} {\bibfnamefont {H.}~\bibnamefont
  {Terrones}},\ }\bibfield  {title} {\bibinfo {title} {Electronic and optical
  properties of strained graphene and other strained {2D} materials: {a}
  review},\ }\href {https://doi.org/10.1088/1361-6633/aa74ef} {\bibfield
  {journal} {\bibinfo  {journal} {Rep. Prog. Phys.}\ }\textbf {\bibinfo
  {volume} {80}},\ \bibinfo {pages} {096501} (\bibinfo {year}
  {2017})}\BibitemShut {NoStop}%
\bibitem [{\citenamefont {de~Juan}\ \emph {et~al.}(2007)\citenamefont
  {de~Juan}, \citenamefont {Cortijo},\ and\ \citenamefont
  {Vozmediano}}]{dejuan_2007}%
  \BibitemOpen
  \bibfield  {author} {\bibinfo {author} {\bibfnamefont {F.}~\bibnamefont
  {de~Juan}}, \bibinfo {author} {\bibfnamefont {A.}~\bibnamefont {Cortijo}},\
  and\ \bibinfo {author} {\bibfnamefont {M.~A.~H.}\ \bibnamefont
  {Vozmediano}},\ }\bibfield  {title} {\bibinfo {title} {Charge inhomogeneities
  due to smooth ripples in graphene sheets},\ }\href
  {https://doi.org/10.1103/PhysRevB.76.165409} {\bibfield  {journal} {\bibinfo
  {journal} {Phys. Rev. B}\ }\textbf {\bibinfo {volume} {76}},\ \bibinfo
  {pages} {165409} (\bibinfo {year} {2007})}\BibitemShut {NoStop}%
\bibitem [{\citenamefont {Gonz{\'a}lez}\ and\ \citenamefont
  {Herrero}(2010)}]{gonzalez_2010}%
  \BibitemOpen
  \bibfield  {author} {\bibinfo {author} {\bibfnamefont {J.}~\bibnamefont
  {Gonz{\'a}lez}}\ and\ \bibinfo {author} {\bibfnamefont {J.}~\bibnamefont
  {Herrero}},\ }\bibfield  {title} {\bibinfo {title} {Graphene wormholes: A
  condensed matter illustration of {D}irac fermions in curved space},\ }\href
  {https://doi.org/https://doi.org/10.1016/j.nuclphysb.2009.09.028} {\bibfield
  {journal} {\bibinfo  {journal} {Nucl. Phys. B}\ }\textbf {\bibinfo {volume}
  {825}},\ \bibinfo {pages} {426} (\bibinfo {year} {2010})}\BibitemShut
  {NoStop}%
\bibitem [{\citenamefont {Stegmann}\ and\ \citenamefont
  {Szpak}(2016)}]{stegmann_2016}%
  \BibitemOpen
  \bibfield  {author} {\bibinfo {author} {\bibfnamefont {T.}~\bibnamefont
  {Stegmann}}\ and\ \bibinfo {author} {\bibfnamefont {N.}~\bibnamefont
  {Szpak}},\ }\bibfield  {title} {\bibinfo {title} {Current flow paths in
  deformed graphene: {f}rom quantum transport to classical trajectories in
  curved space},\ }\href {https://doi.org/10.1088/1367-2630/18/5/053016}
  {\bibfield  {journal} {\bibinfo  {journal} {New J. Phys.}\ }\textbf {\bibinfo
  {volume} {18}},\ \bibinfo {pages} {053016} (\bibinfo {year}
  {2016})}\BibitemShut {NoStop}%
\bibitem [{\citenamefont {Castro-Villarreal}\ and\ \citenamefont
  {Ruiz-S\'anchez}(2017)}]{castro-villareal_2017}%
  \BibitemOpen
  \bibfield  {author} {\bibinfo {author} {\bibfnamefont {P.}~\bibnamefont
  {Castro-Villarreal}}\ and\ \bibinfo {author} {\bibfnamefont {R.}~\bibnamefont
  {Ruiz-S\'anchez}},\ }\bibfield  {title} {\bibinfo {title} {Pseudomagnetic
  field in curved graphene},\ }\href
  {https://doi.org/10.1103/PhysRevB.95.125432} {\bibfield  {journal} {\bibinfo
  {journal} {Phys. Rev. B}\ }\textbf {\bibinfo {volume} {95}},\ \bibinfo
  {pages} {125432} (\bibinfo {year} {2017})}\BibitemShut {NoStop}%
\bibitem [{\citenamefont {Gallerati}(2021)}]{gallerati_2021}%
  \BibitemOpen
  \bibfield  {author} {\bibinfo {author} {\bibfnamefont {A.}~\bibnamefont
  {Gallerati}},\ }\bibfield  {title} {\bibinfo {title} {Negative-curvature
  spacetime solutions for graphene},\ }\href
  {https://doi.org/10.1088/1361-648X/abd9a2} {\bibfield  {journal} {\bibinfo
  {journal} {J. Phys. Condens. Matter}\ }\textbf {\bibinfo {volume} {33}},\
  \bibinfo {pages} {135501} (\bibinfo {year} {2021})}\BibitemShut {NoStop}%
\bibitem [{\citenamefont {Roche}\ \emph {et~al.}(2007)\citenamefont {Roche},
  \citenamefont {Jiang}, \citenamefont {Torres},\ and\ \citenamefont
  {Saito}}]{roche_2007}%
  \BibitemOpen
  \bibfield  {author} {\bibinfo {author} {\bibfnamefont {S.}~\bibnamefont
  {Roche}}, \bibinfo {author} {\bibfnamefont {J.}~\bibnamefont {Jiang}},
  \bibinfo {author} {\bibfnamefont {L.~E. F.~F.}\ \bibnamefont {Torres}},\ and\
  \bibinfo {author} {\bibfnamefont {R.}~\bibnamefont {Saito}},\ }\bibfield
  {title} {\bibinfo {title} {Charge transport in carbon nanotubes: {q}uantum
  effects of electron--phonon coupling},\ }\href
  {https://doi.org/10.1088/0953-8984/19/18/183203} {\bibfield  {journal}
  {\bibinfo  {journal} {J. Phys. Condens. Matter}\ }\textbf {\bibinfo {volume}
  {19}},\ \bibinfo {pages} {183203} (\bibinfo {year} {2007})}\BibitemShut
  {NoStop}%
\bibitem [{\citenamefont {Charlier}\ \emph {et~al.}(2007)\citenamefont
  {Charlier}, \citenamefont {Blase},\ and\ \citenamefont
  {Roche}}]{charlier_2007}%
  \BibitemOpen
  \bibfield  {author} {\bibinfo {author} {\bibfnamefont {J.-C.}\ \bibnamefont
  {Charlier}}, \bibinfo {author} {\bibfnamefont {X.}~\bibnamefont {Blase}},\
  and\ \bibinfo {author} {\bibfnamefont {S.}~\bibnamefont {Roche}},\ }\bibfield
   {title} {\bibinfo {title} {Electronic and transport properties of
  nanotubes},\ }\href {https://doi.org/10.1103/RevModPhys.79.677} {\bibfield
  {journal} {\bibinfo  {journal} {Rev. Mod. Phys.}\ }\textbf {\bibinfo {volume}
  {79}},\ \bibinfo {pages} {677} (\bibinfo {year} {2007})}\BibitemShut
  {NoStop}%
\bibitem [{\citenamefont {Dubois}\ \emph {et~al.}(2009)\citenamefont {Dubois},
  \citenamefont {Zanolli}, \citenamefont {Declerck},\ and\ \citenamefont
  {Charlier}}]{dubois_2009}%
  \BibitemOpen
  \bibfield  {author} {\bibinfo {author} {\bibfnamefont {S.~M.~M.}\
  \bibnamefont {Dubois}}, \bibinfo {author} {\bibfnamefont {Z.}~\bibnamefont
  {Zanolli}}, \bibinfo {author} {\bibfnamefont {X.}~\bibnamefont {Declerck}},\
  and\ \bibinfo {author} {\bibfnamefont {J.~C.}\ \bibnamefont {Charlier}},\
  }\bibfield  {title} {\bibinfo {title} {Electronic properties and quantum
  transport in {G}raphene-based nanostructures},\ }\href
  {https://doi.org/10.1140/epjb/e2009-00327-8} {\bibfield  {journal} {\bibinfo
  {journal} {Eur. Phys. J. B}\ }\textbf {\bibinfo {volume} {72}},\ \bibinfo
  {pages} {1} (\bibinfo {year} {2009})}\BibitemShut {NoStop}%
\bibitem [{\citenamefont {Laird}\ \emph {et~al.}(2015)\citenamefont {Laird},
  \citenamefont {Kuemmeth}, \citenamefont {Steele}, \citenamefont
  {Grove-Rasmussen}, \citenamefont {Nyg\aa{}rd}, \citenamefont {Flensberg},\
  and\ \citenamefont {Kouwenhoven}}]{laird_2015}%
  \BibitemOpen
  \bibfield  {author} {\bibinfo {author} {\bibfnamefont {E.~A.}\ \bibnamefont
  {Laird}}, \bibinfo {author} {\bibfnamefont {F.}~\bibnamefont {Kuemmeth}},
  \bibinfo {author} {\bibfnamefont {G.~A.}\ \bibnamefont {Steele}}, \bibinfo
  {author} {\bibfnamefont {K.}~\bibnamefont {Grove-Rasmussen}}, \bibinfo
  {author} {\bibfnamefont {J.}~\bibnamefont {Nyg\aa{}rd}}, \bibinfo {author}
  {\bibfnamefont {K.}~\bibnamefont {Flensberg}},\ and\ \bibinfo {author}
  {\bibfnamefont {L.~P.}\ \bibnamefont {Kouwenhoven}},\ }\bibfield  {title}
  {\bibinfo {title} {Quantum transport in carbon nanotubes},\ }\href
  {https://doi.org/10.1103/RevModPhys.87.703} {\bibfield  {journal} {\bibinfo
  {journal} {Rev. Mod. Phys.}\ }\textbf {\bibinfo {volume} {87}},\ \bibinfo
  {pages} {703} (\bibinfo {year} {2015})}\BibitemShut {NoStop}%
\bibitem [{\citenamefont {Iijima}(1991)}]{iijima_1991}%
  \BibitemOpen
  \bibfield  {author} {\bibinfo {author} {\bibfnamefont {S.}~\bibnamefont
  {Iijima}},\ }\bibfield  {title} {\bibinfo {title} {Helical microtubules of
  graphitic carbon},\ }\href {https://doi.org/10.1038/354056a0} {\bibfield
  {journal} {\bibinfo  {journal} {Nature}\ }\textbf {\bibinfo {volume} {354}},\
  \bibinfo {pages} {56} (\bibinfo {year} {1991})}\BibitemShut {NoStop}%
\bibitem [{\citenamefont {Saito}\ \emph {et~al.}(1998)\citenamefont {Saito},
  \citenamefont {Dresselhaus},\ and\ \citenamefont {Dresselhaus}}]{Saito1998}%
  \BibitemOpen
  \bibfield  {author} {\bibinfo {author} {\bibfnamefont {R.}~\bibnamefont
  {Saito}}, \bibinfo {author} {\bibfnamefont {G.}~\bibnamefont {Dresselhaus}},\
  and\ \bibinfo {author} {\bibfnamefont {M.~S.}\ \bibnamefont {Dresselhaus}},\
  }\href {https://doi.org/10.1142/p080} {\emph {\bibinfo {title} {Physical
  Properties of Carbon Nanotubes}}}\ (\bibinfo  {publisher} {Imperial College
  Press, London},\ \bibinfo {year} {1998})\BibitemShut {NoStop}%
\bibitem [{\citenamefont {Reich}\ \emph {et~al.}(2004)\citenamefont {Reich},
  \citenamefont {Thomsen},\ and\ \citenamefont {Maultzsch}}]{Reich2004}%
  \BibitemOpen
  \bibfield  {author} {\bibinfo {author} {\bibfnamefont {S.}~\bibnamefont
  {Reich}}, \bibinfo {author} {\bibfnamefont {C.}~\bibnamefont {Thomsen}},\
  and\ \bibinfo {author} {\bibfnamefont {J.}~\bibnamefont {Maultzsch}},\ }\href
  {https://doi.org/10.1002/9783527618040} {\emph {\bibinfo {title} {Carbon
  Nanotubes}}}\ (\bibinfo  {publisher} {Wiley-VCH, Berlin},\ \bibinfo {year}
  {2004})\BibitemShut {NoStop}%
\bibitem [{\citenamefont {Lassagne}\ \emph {et~al.}(2009)\citenamefont
  {Lassagne}, \citenamefont {Tarakanov}, \citenamefont {Kinaret}, \citenamefont
  {Garcia-Sanchez},\ and\ \citenamefont {Bachtold}}]{lassagne_2009}%
  \BibitemOpen
  \bibfield  {author} {\bibinfo {author} {\bibfnamefont {B.}~\bibnamefont
  {Lassagne}}, \bibinfo {author} {\bibfnamefont {Y.}~\bibnamefont {Tarakanov}},
  \bibinfo {author} {\bibfnamefont {J.}~\bibnamefont {Kinaret}}, \bibinfo
  {author} {\bibfnamefont {D.}~\bibnamefont {Garcia-Sanchez}},\ and\ \bibinfo
  {author} {\bibfnamefont {A.}~\bibnamefont {Bachtold}},\ }\bibfield  {title}
  {\bibinfo {title} {{Coupling Mechanics to Charge Transport in Carbon Nanotube
  Mechanical Resonators}},\ }\href {https://doi.org/10.1126/science.1174290}
  {\bibfield  {journal} {\bibinfo  {journal} {Science}\ }\textbf {\bibinfo
  {volume} {325}},\ \bibinfo {pages} {1107} (\bibinfo {year}
  {2009})}\BibitemShut {NoStop}%
\bibitem [{\citenamefont {Eichler}\ \emph {et~al.}(2011)\citenamefont
  {Eichler}, \citenamefont {Moser}, \citenamefont {Chaste}, \citenamefont
  {Zdrojek}, \citenamefont {Wilson-Rae},\ and\ \citenamefont
  {Bachtold}}]{eichler_2011}%
  \BibitemOpen
  \bibfield  {author} {\bibinfo {author} {\bibfnamefont {A.}~\bibnamefont
  {Eichler}}, \bibinfo {author} {\bibfnamefont {J.}~\bibnamefont {Moser}},
  \bibinfo {author} {\bibfnamefont {J.}~\bibnamefont {Chaste}}, \bibinfo
  {author} {\bibfnamefont {M.}~\bibnamefont {Zdrojek}}, \bibinfo {author}
  {\bibfnamefont {I.}~\bibnamefont {Wilson-Rae}},\ and\ \bibinfo {author}
  {\bibfnamefont {A.}~\bibnamefont {Bachtold}},\ }\bibfield  {title} {\bibinfo
  {title} {Nonlinear damping in mechanical resonators made from carbon
  nanotubes and graphene},\ }\href {https://doi.org/10.1038/nnano.2011.71}
  {\bibfield  {journal} {\bibinfo  {journal} {Nat. Nanotechnol.}\ }\textbf
  {\bibinfo {volume} {6}},\ \bibinfo {pages} {339} (\bibinfo {year}
  {2011})}\BibitemShut {NoStop}%
\bibitem [{\citenamefont {P\'alyi}\ \emph {et~al.}(2012)\citenamefont
  {P\'alyi}, \citenamefont {Struck}, \citenamefont {Rudner}, \citenamefont
  {Flensberg},\ and\ \citenamefont {Burkard}}]{palyi_2012}%
  \BibitemOpen
  \bibfield  {author} {\bibinfo {author} {\bibfnamefont {A.}~\bibnamefont
  {P\'alyi}}, \bibinfo {author} {\bibfnamefont {P.~R.}\ \bibnamefont {Struck}},
  \bibinfo {author} {\bibfnamefont {M.}~\bibnamefont {Rudner}}, \bibinfo
  {author} {\bibfnamefont {K.}~\bibnamefont {Flensberg}},\ and\ \bibinfo
  {author} {\bibfnamefont {G.}~\bibnamefont {Burkard}},\ }\bibfield  {title}
  {\bibinfo {title} {{Spin-Orbit-Induced Strong Coupling of a Single Spin to a
  Nanomechanical Resonator}},\ }\href
  {https://doi.org/10.1103/PhysRevLett.108.206811} {\bibfield  {journal}
  {\bibinfo  {journal} {Phys. Rev. Lett.}\ }\textbf {\bibinfo {volume} {108}},\
  \bibinfo {pages} {206811} (\bibinfo {year} {2012})}\BibitemShut {NoStop}%
\bibitem [{\citenamefont {Wang}\ and\ \citenamefont
  {Burkard}(2016)}]{wang_2016}%
  \BibitemOpen
  \bibfield  {author} {\bibinfo {author} {\bibfnamefont {H.}~\bibnamefont
  {Wang}}\ and\ \bibinfo {author} {\bibfnamefont {G.}~\bibnamefont {Burkard}},\
  }\bibfield  {title} {\bibinfo {title} {Creating arbitrary quantum vibrational
  states in a carbon nanotube},\ }\href
  {https://doi.org/10.1103/PhysRevB.94.205413} {\bibfield  {journal} {\bibinfo
  {journal} {Phys. Rev. B}\ }\textbf {\bibinfo {volume} {94}},\ \bibinfo
  {pages} {205413} (\bibinfo {year} {2016})}\BibitemShut {NoStop}%
\bibitem [{\citenamefont {Kane}\ and\ \citenamefont {Mele}(1997)}]{kane_1997}%
  \BibitemOpen
  \bibfield  {author} {\bibinfo {author} {\bibfnamefont {C.~L.}\ \bibnamefont
  {Kane}}\ and\ \bibinfo {author} {\bibfnamefont {E.~J.}\ \bibnamefont
  {Mele}},\ }\bibfield  {title} {\bibinfo {title} {{Size, Shape, and Low Energy
  Electronic Structure of Carbon Nanotubes}},\ }\href
  {https://doi.org/10.1103/PhysRevLett.78.1932} {\bibfield  {journal} {\bibinfo
   {journal} {Phys. Rev. Lett.}\ }\textbf {\bibinfo {volume} {78}},\ \bibinfo
  {pages} {1932} (\bibinfo {year} {1997})}\BibitemShut {NoStop}%
\bibitem [{\citenamefont {Rochefort}\ \emph {et~al.}(1999)\citenamefont
  {Rochefort}, \citenamefont {Avouris}, \citenamefont {Lesage},\ and\
  \citenamefont {Salahub}}]{rochefort_1999}%
  \BibitemOpen
  \bibfield  {author} {\bibinfo {author} {\bibfnamefont {A.}~\bibnamefont
  {Rochefort}}, \bibinfo {author} {\bibfnamefont {P.}~\bibnamefont {Avouris}},
  \bibinfo {author} {\bibfnamefont {F.}~\bibnamefont {Lesage}},\ and\ \bibinfo
  {author} {\bibfnamefont {D.~R.}\ \bibnamefont {Salahub}},\ }\bibfield
  {title} {\bibinfo {title} {Electrical and mechanical properties of distorted
  carbon nanotubes},\ }\href {https://doi.org/10.1103/PhysRevB.60.13824}
  {\bibfield  {journal} {\bibinfo  {journal} {Phys. Rev. B}\ }\textbf {\bibinfo
  {volume} {60}},\ \bibinfo {pages} {13824} (\bibinfo {year}
  {1999})}\BibitemShut {NoStop}%
\bibitem [{\citenamefont {Tombler}\ \emph {et~al.}(2000)\citenamefont
  {Tombler}, \citenamefont {Zhou}, \citenamefont {Alexseyev}, \citenamefont
  {Kong}, \citenamefont {Dai}, \citenamefont {Liu}, \citenamefont {Jayanthi},
  \citenamefont {Tang},\ and\ \citenamefont {Wu}}]{Tombler2000}%
  \BibitemOpen
  \bibfield  {author} {\bibinfo {author} {\bibfnamefont {T.~W.}\ \bibnamefont
  {Tombler}}, \bibinfo {author} {\bibfnamefont {C.}~\bibnamefont {Zhou}},
  \bibinfo {author} {\bibfnamefont {L.}~\bibnamefont {Alexseyev}}, \bibinfo
  {author} {\bibfnamefont {J.}~\bibnamefont {Kong}}, \bibinfo {author}
  {\bibfnamefont {H.}~\bibnamefont {Dai}}, \bibinfo {author} {\bibfnamefont
  {L.}~\bibnamefont {Liu}}, \bibinfo {author} {\bibfnamefont {C.~S.}\
  \bibnamefont {Jayanthi}}, \bibinfo {author} {\bibfnamefont {M.}~\bibnamefont
  {Tang}},\ and\ \bibinfo {author} {\bibfnamefont {S.-Y.}\ \bibnamefont {Wu}},\
  }\bibfield  {title} {\bibinfo {title} {Reversible electromechanical
  characteristics of carbon nanotubes underlocal-probe manipulation},\ }\href
  {https://doi.org/10.1038/35015519} {\bibfield  {journal} {\bibinfo  {journal}
  {Nature}\ }\textbf {\bibinfo {volume} {405}},\ \bibinfo {pages} {769}
  (\bibinfo {year} {2000})}\BibitemShut {NoStop}%
\bibitem [{\citenamefont {Suzuura}\ and\ \citenamefont
  {Ando}(2002)}]{suzuura_2002}%
  \BibitemOpen
  \bibfield  {author} {\bibinfo {author} {\bibfnamefont {H.}~\bibnamefont
  {Suzuura}}\ and\ \bibinfo {author} {\bibfnamefont {T.}~\bibnamefont {Ando}},\
  }\bibfield  {title} {\bibinfo {title} {Phonons and electron-phonon scattering
  in carbon nanotubes},\ }\href {https://doi.org/10.1103/PhysRevB.65.235412}
  {\bibfield  {journal} {\bibinfo  {journal} {Phys. Rev. B}\ }\textbf {\bibinfo
  {volume} {65}},\ \bibinfo {pages} {235412} (\bibinfo {year}
  {2002})}\BibitemShut {NoStop}%
\bibitem [{\citenamefont {Farajian}\ \emph {et~al.}(2003)\citenamefont
  {Farajian}, \citenamefont {Yakobson}, \citenamefont {Mizuseki},\ and\
  \citenamefont {Kawazoe}}]{farajian_2003}%
  \BibitemOpen
  \bibfield  {author} {\bibinfo {author} {\bibfnamefont {A.~A.}\ \bibnamefont
  {Farajian}}, \bibinfo {author} {\bibfnamefont {B.~I.}\ \bibnamefont
  {Yakobson}}, \bibinfo {author} {\bibfnamefont {H.}~\bibnamefont {Mizuseki}},\
  and\ \bibinfo {author} {\bibfnamefont {Y.}~\bibnamefont {Kawazoe}},\
  }\bibfield  {title} {\bibinfo {title} {Electronic transport through bent
  carbon nanotubes: Nanoelectromechanical sensors and switches},\ }\href
  {https://doi.org/10.1103/PhysRevB.67.205423} {\bibfield  {journal} {\bibinfo
  {journal} {Phys. Rev. B}\ }\textbf {\bibinfo {volume} {67}},\ \bibinfo
  {pages} {205423} (\bibinfo {year} {2003})}\BibitemShut {NoStop}%
\bibitem [{\citenamefont {Fa}\ and\ \citenamefont {Dong}(2004)}]{fa_2004}%
  \BibitemOpen
  \bibfield  {author} {\bibinfo {author} {\bibfnamefont {W.}~\bibnamefont
  {Fa}}\ and\ \bibinfo {author} {\bibfnamefont {J.}~\bibnamefont {Dong}},\
  }\bibfield  {title} {\bibinfo {title} {Quantum interference in deformed
  carbon nanotube waveguides},\ }\href
  {https://doi.org/10.1103/PhysRevB.70.233407} {\bibfield  {journal} {\bibinfo
  {journal} {Phys. Rev. B}\ }\textbf {\bibinfo {volume} {70}},\ \bibinfo
  {pages} {233407} (\bibinfo {year} {2004})}\BibitemShut {NoStop}%
\bibitem [{\citenamefont {Koskinen}(2010)}]{koskinen_2010}%
  \BibitemOpen
  \bibfield  {author} {\bibinfo {author} {\bibfnamefont {P.}~\bibnamefont
  {Koskinen}},\ }\bibfield  {title} {\bibinfo {title} {Electronic and optical
  properties of carbon nanotubes under pure bending},\ }\href
  {https://doi.org/10.1103/PhysRevB.82.193409} {\bibfield  {journal} {\bibinfo
  {journal} {Phys. Rev. B}\ }\textbf {\bibinfo {volume} {82}},\ \bibinfo
  {pages} {193409} (\bibinfo {year} {2010})}\BibitemShut {NoStop}%
\bibitem [{\citenamefont {Wang}\ \emph {et~al.}(2010)\citenamefont {Wang},
  \citenamefont {Gupta}, \citenamefont {Huang}, \citenamefont {Vedala},
  \citenamefont {Hao}, \citenamefont {Crespi}, \citenamefont {Choi},\ and\
  \citenamefont {Eklund}}]{wang_2010}%
  \BibitemOpen
  \bibfield  {author} {\bibinfo {author} {\bibfnamefont {B.}~\bibnamefont
  {Wang}}, \bibinfo {author} {\bibfnamefont {A.~K.}\ \bibnamefont {Gupta}},
  \bibinfo {author} {\bibfnamefont {J.}~\bibnamefont {Huang}}, \bibinfo
  {author} {\bibfnamefont {H.}~\bibnamefont {Vedala}}, \bibinfo {author}
  {\bibfnamefont {Q.}~\bibnamefont {Hao}}, \bibinfo {author} {\bibfnamefont
  {V.~H.}\ \bibnamefont {Crespi}}, \bibinfo {author} {\bibfnamefont
  {W.}~\bibnamefont {Choi}},\ and\ \bibinfo {author} {\bibfnamefont {P.~C.}\
  \bibnamefont {Eklund}},\ }\bibfield  {title} {\bibinfo {title} {Effect of
  bending on single-walled carbon nanotubes: A {R}aman scattering study},\
  }\href {https://doi.org/10.1103/PhysRevB.81.115422} {\bibfield  {journal}
  {\bibinfo  {journal} {Phys. Rev. B}\ }\textbf {\bibinfo {volume} {81}},\
  \bibinfo {pages} {115422} (\bibinfo {year} {2010})}\BibitemShut {NoStop}%
\bibitem [{\citenamefont {Shima}(2012)}]{shima_2012}%
  \BibitemOpen
  \bibfield  {author} {\bibinfo {author} {\bibfnamefont {H.}~\bibnamefont
  {Shima}},\ }\bibfield  {title} {\bibinfo {title} {{Buckling of Carbon
  Nanotubes: A State of the Art Review}},\ }\href
  {https://doi.org/10.3390/ma5010047} {\bibfield  {journal} {\bibinfo
  {journal} {Materials}\ }\textbf {\bibinfo {volume} {5}},\ \bibinfo {pages}
  {47} (\bibinfo {year} {2012})}\BibitemShut {NoStop}%
\bibitem [{\citenamefont {Rahman}\ \emph {et~al.}(2017)\citenamefont {Rahman},
  \citenamefont {Chowdhury}, \citenamefont {Rosul},\ and\ \citenamefont
  {Alam}}]{rahman_2017}%
  \BibitemOpen
  \bibfield  {author} {\bibinfo {author} {\bibfnamefont {M.~M.}\ \bibnamefont
  {Rahman}}, \bibinfo {author} {\bibfnamefont {M.~M.}\ \bibnamefont
  {Chowdhury}}, \bibinfo {author} {\bibfnamefont {M.~G.}\ \bibnamefont
  {Rosul}},\ and\ \bibinfo {author} {\bibfnamefont {M.~K.}\ \bibnamefont
  {Alam}},\ }\bibfield  {title} {\bibinfo {title} {Effect of bending on the
  molecular transport along carbon nanotubes},\ }\href
  {https://doi.org/https://doi.org/10.1002/pssb.201600266} {\bibfield
  {journal} {\bibinfo  {journal} {Phys. Status Solidi B}\ }\textbf {\bibinfo
  {volume} {254}},\ \bibinfo {pages} {1600266} (\bibinfo {year}
  {2017})}\BibitemShut {NoStop}%
\bibitem [{\citenamefont {Wu}\ \emph {et~al.}(2019)\citenamefont {Wu},
  \citenamefont {Xing}, \citenamefont {Ren}, \citenamefont {Wang},\ and\
  \citenamefont {Guo}}]{wu_2019}%
  \BibitemOpen
  \bibfield  {author} {\bibinfo {author} {\bibfnamefont {Z.}~\bibnamefont
  {Wu}}, \bibinfo {author} {\bibfnamefont {Y.}~\bibnamefont {Xing}}, \bibinfo
  {author} {\bibfnamefont {W.}~\bibnamefont {Ren}}, \bibinfo {author}
  {\bibfnamefont {Y.}~\bibnamefont {Wang}},\ and\ \bibinfo {author}
  {\bibfnamefont {H.}~\bibnamefont {Guo}},\ }\bibfield  {title} {\bibinfo
  {title} {Ballistic transport in bent-shaped carbon nanotubes},\ }\href
  {https://doi.org/https://doi.org/10.1016/j.carbon.2019.04.062} {\bibfield
  {journal} {\bibinfo  {journal} {Carbon}\ }\textbf {\bibinfo {volume} {149}},\
  \bibinfo {pages} {364} (\bibinfo {year} {2019})}\BibitemShut {NoStop}%
\bibitem [{\citenamefont {Ceulemans}\ \emph {et~al.}(2000)\citenamefont
  {Ceulemans}, \citenamefont {Chibotaru}, \citenamefont {Bovin},\ and\
  \citenamefont {Fowler}}]{ceulemans_2000}%
  \BibitemOpen
  \bibfield  {author} {\bibinfo {author} {\bibfnamefont {A.}~\bibnamefont
  {Ceulemans}}, \bibinfo {author} {\bibfnamefont {L.~F.}\ \bibnamefont
  {Chibotaru}}, \bibinfo {author} {\bibfnamefont {S.~A.}\ \bibnamefont
  {Bovin}},\ and\ \bibinfo {author} {\bibfnamefont {P.~W.}\ \bibnamefont
  {Fowler}},\ }\bibfield  {title} {\bibinfo {title} {The electronic structure
  of polyhex carbon tori},\ }\href {https://doi.org/10.1063/1.480972}
  {\bibfield  {journal} {\bibinfo  {journal} {J. Chem. Phys.}\ }\textbf
  {\bibinfo {volume} {112}},\ \bibinfo {pages} {4271} (\bibinfo {year}
  {2000})},\ \Eprint {https://arxiv.org/abs/https://doi.org/10.1063/1.480972}
  {https://doi.org/10.1063/1.480972} \BibitemShut {NoStop}%
\bibitem [{\citenamefont {Zhang}\ \emph {et~al.}(2005)\citenamefont {Zhang},
  \citenamefont {Yang}, \citenamefont {Wang}, \citenamefont {Yuan},
  \citenamefont {Zhang}, \citenamefont {Qiu},\ and\ \citenamefont
  {Peng}}]{Zhang_2005}%
  \BibitemOpen
  \bibfield  {author} {\bibinfo {author} {\bibfnamefont {Z.}~\bibnamefont
  {Zhang}}, \bibinfo {author} {\bibfnamefont {Z.}~\bibnamefont {Yang}},
  \bibinfo {author} {\bibfnamefont {X.}~\bibnamefont {Wang}}, \bibinfo {author}
  {\bibfnamefont {J.}~\bibnamefont {Yuan}}, \bibinfo {author} {\bibfnamefont
  {H.}~\bibnamefont {Zhang}}, \bibinfo {author} {\bibfnamefont
  {M.}~\bibnamefont {Qiu}},\ and\ \bibinfo {author} {\bibfnamefont
  {J.}~\bibnamefont {Peng}},\ }\bibfield  {title} {\bibinfo {title} {The
  electronic structure of a deformed chiral carbon nanotorus},\ }\href
  {https://doi.org/10.1088/0953-8984/17/26/010} {\bibfield  {journal} {\bibinfo
   {journal} {J. Phys. Condens.}\ }\textbf {\bibinfo {volume} {17}},\ \bibinfo
  {pages} {4111} (\bibinfo {year} {2005})}\BibitemShut {NoStop}%
\bibitem [{\citenamefont {Liu}\ \emph {et~al.}(2014)\citenamefont {Liu},
  \citenamefont {Liu},\ and\ \citenamefont {Zhao}}]{Lizhao_2014}%
  \BibitemOpen
  \bibfield  {author} {\bibinfo {author} {\bibfnamefont {L.}~\bibnamefont
  {Liu}}, \bibinfo {author} {\bibfnamefont {F.}~\bibnamefont {Liu}},\ and\
  \bibinfo {author} {\bibfnamefont {J.}~\bibnamefont {Zhao}},\ }\bibfield
  {title} {\bibinfo {title} {{Curved carbon nanotubes: From unique geometries
  to novel properties and peculiar applications}},\ }\href
  {https://doi.org/10.1007/s12274-014-0431-1} {\bibfield  {journal} {\bibinfo
  {journal} {Nano Res.}\ }\textbf {\bibinfo {volume} {7}},\ \bibinfo {pages}
  {626} (\bibinfo {year} {2014})}\BibitemShut {NoStop}%
\bibitem [{\citenamefont {Chaves}\ \emph {et~al.}(2010)\citenamefont {Chaves},
  \citenamefont {Covaci}, \citenamefont {Rakhimov}, \citenamefont {Farias},\
  and\ \citenamefont {Peeters}}]{chaves_2010}%
  \BibitemOpen
  \bibfield  {author} {\bibinfo {author} {\bibfnamefont {A.}~\bibnamefont
  {Chaves}}, \bibinfo {author} {\bibfnamefont {L.}~\bibnamefont {Covaci}},
  \bibinfo {author} {\bibfnamefont {K.~Y.}\ \bibnamefont {Rakhimov}}, \bibinfo
  {author} {\bibfnamefont {G.~A.}\ \bibnamefont {Farias}},\ and\ \bibinfo
  {author} {\bibfnamefont {F.~M.}\ \bibnamefont {Peeters}},\ }\bibfield
  {title} {\bibinfo {title} {Wave-packet dynamics and valley filter in strained
  graphene},\ }\href {https://doi.org/10.1103/PhysRevB.82.205430} {\bibfield
  {journal} {\bibinfo  {journal} {Phys. Rev. B}\ }\textbf {\bibinfo {volume}
  {82}},\ \bibinfo {pages} {205430} (\bibinfo {year} {2010})}\BibitemShut
  {NoStop}%
\bibitem [{\citenamefont {Settnes}\ \emph {et~al.}(2016)\citenamefont
  {Settnes}, \citenamefont {Power}, \citenamefont {Brandbyge},\ and\
  \citenamefont {Jauho}}]{Settnes2016}%
  \BibitemOpen
  \bibfield  {author} {\bibinfo {author} {\bibfnamefont {M.}~\bibnamefont
  {Settnes}}, \bibinfo {author} {\bibfnamefont {S.~R.}\ \bibnamefont {Power}},
  \bibinfo {author} {\bibfnamefont {M.}~\bibnamefont {Brandbyge}},\ and\
  \bibinfo {author} {\bibfnamefont {A.-P.}\ \bibnamefont {Jauho}},\ }\bibfield
  {title} {\bibinfo {title} {{Graphene Nanobubbles as Valley Filters and Beam
  Splitters}},\ }\href {https://doi.org/10.1103/PhysRevLett.117.276801}
  {\bibfield  {journal} {\bibinfo  {journal} {Phys. Rev. Lett.}\ }\textbf
  {\bibinfo {volume} {117}},\ \bibinfo {pages} {276801} (\bibinfo {year}
  {2016})}\BibitemShut {NoStop}%
\bibitem [{\citenamefont {Milovanovi{\'c}}\ and\ \citenamefont
  {Peeters}(2016)}]{Milovanovic2016}%
  \BibitemOpen
  \bibfield  {author} {\bibinfo {author} {\bibfnamefont {S.~P.}\ \bibnamefont
  {Milovanovi{\'c}}}\ and\ \bibinfo {author} {\bibfnamefont {F.~M.}\
  \bibnamefont {Peeters}},\ }\bibfield  {title} {\bibinfo {title} {Strain
  controlled valley filtering in multi-terminal graphene structures},\ }\href
  {https://doi.org/10.1063/1.4967977} {\bibfield  {journal} {\bibinfo
  {journal} {Appl. Phys. Lett.}\ }\textbf {\bibinfo {volume} {109}},\ \bibinfo
  {pages} {203108} (\bibinfo {year} {2016})}\BibitemShut {NoStop}%
\bibitem [{\citenamefont {Schaibley}\ \emph {et~al.}(2016)\citenamefont
  {Schaibley}, \citenamefont {Yu}, \citenamefont {Clark}, \citenamefont
  {Rivera}, \citenamefont {Ross}, \citenamefont {Seyler}, \citenamefont {Yao},\
  and\ \citenamefont {Xu}}]{Schaibley2016}%
  \BibitemOpen
  \bibfield  {author} {\bibinfo {author} {\bibfnamefont {J.~R.}\ \bibnamefont
  {Schaibley}}, \bibinfo {author} {\bibfnamefont {H.}~\bibnamefont {Yu}},
  \bibinfo {author} {\bibfnamefont {G.}~\bibnamefont {Clark}}, \bibinfo
  {author} {\bibfnamefont {P.}~\bibnamefont {Rivera}}, \bibinfo {author}
  {\bibfnamefont {J.~S.}\ \bibnamefont {Ross}}, \bibinfo {author}
  {\bibfnamefont {K.~L.}\ \bibnamefont {Seyler}}, \bibinfo {author}
  {\bibfnamefont {W.}~\bibnamefont {Yao}},\ and\ \bibinfo {author}
  {\bibfnamefont {X.}~\bibnamefont {Xu}},\ }\bibfield  {title} {\bibinfo
  {title} {Valleytronics in {2D} materials},\ }\href
  {https://doi.org/10.1038/natrevmats.2016.55} {\bibfield  {journal} {\bibinfo
  {journal} {Nat. Rev. Mater.}\ }\textbf {\bibinfo {volume} {1}},\ \bibinfo
  {pages} {16055} (\bibinfo {year} {2016})}\BibitemShut {NoStop}%
\bibitem [{\citenamefont {Osika}\ \emph {et~al.}(2017)\citenamefont {Osika},
  \citenamefont {Chac{\'o}n}, \citenamefont {Lewenstein},\ and\ \citenamefont
  {Szafran}}]{osika_2017}%
  \BibitemOpen
  \bibfield  {author} {\bibinfo {author} {\bibfnamefont {E.~N.}\ \bibnamefont
  {Osika}}, \bibinfo {author} {\bibfnamefont {A.}~\bibnamefont {Chac{\'o}n}},
  \bibinfo {author} {\bibfnamefont {M.}~\bibnamefont {Lewenstein}},\ and\
  \bibinfo {author} {\bibfnamefont {B.}~\bibnamefont {Szafran}},\ }\bibfield
  {title} {\bibinfo {title} {Spin-valley dynamics of electrically driven
  ambipolar carbon-nanotube quantum dots},\ }\href
  {https://doi.org/10.1088/1361-648X/aa720e} {\bibfield  {journal} {\bibinfo
  {journal} {J. Phys. Condens. Matter}\ }\textbf {\bibinfo {volume} {29}},\
  \bibinfo {pages} {285301} (\bibinfo {year} {2017})}\BibitemShut {NoStop}%
\bibitem [{\citenamefont {Carrillo-Bastos}\ \emph {et~al.}(2018)\citenamefont
  {Carrillo-Bastos}, \citenamefont {Ochoa}, \citenamefont {Zavala},\ and\
  \citenamefont {Mireles}}]{CarrilloBastos2018}%
  \BibitemOpen
  \bibfield  {author} {\bibinfo {author} {\bibfnamefont {R.}~\bibnamefont
  {Carrillo-Bastos}}, \bibinfo {author} {\bibfnamefont {M.}~\bibnamefont
  {Ochoa}}, \bibinfo {author} {\bibfnamefont {S.~A.}\ \bibnamefont {Zavala}},\
  and\ \bibinfo {author} {\bibfnamefont {F.}~\bibnamefont {Mireles}},\
  }\bibfield  {title} {\bibinfo {title} {Enhanced asymmetric valley scattering
  by scalar fields in nonuniform out-of-plane deformations in graphene},\
  }\href {https://doi.org/10.1103/PhysRevB.98.165436} {\bibfield  {journal}
  {\bibinfo  {journal} {Phys. Rev. B}\ }\textbf {\bibinfo {volume} {98}},\
  \bibinfo {pages} {165436} (\bibinfo {year} {2018})}\BibitemShut {NoStop}%
\bibitem [{\citenamefont {Zhai}\ and\ \citenamefont
  {Sandler}(2018)}]{Sandler2018}%
  \BibitemOpen
  \bibfield  {author} {\bibinfo {author} {\bibfnamefont {D.}~\bibnamefont
  {Zhai}}\ and\ \bibinfo {author} {\bibfnamefont {N.}~\bibnamefont {Sandler}},\
  }\bibfield  {title} {\bibinfo {title} {Local versus extended deformed
  graphene geometries for valley filtering},\ }\href
  {https://doi.org/10.1103/PhysRevB.98.165437} {\bibfield  {journal} {\bibinfo
  {journal} {Phys. Rev. B}\ }\textbf {\bibinfo {volume} {98}},\ \bibinfo
  {pages} {165437} (\bibinfo {year} {2018})}\BibitemShut {NoStop}%
\bibitem [{\citenamefont {Stegmann}\ and\ \citenamefont
  {Szpak}(2018)}]{Stegmann2018}%
  \BibitemOpen
  \bibfield  {author} {\bibinfo {author} {\bibfnamefont {T.}~\bibnamefont
  {Stegmann}}\ and\ \bibinfo {author} {\bibfnamefont {N.}~\bibnamefont
  {Szpak}},\ }\bibfield  {title} {\bibinfo {title} {Current splitting and
  valley polarization in elastically deformed graphene},\ }\href
  {https://doi.org/10.1088/2053-1583/aaea8d} {\bibfield  {journal} {\bibinfo
  {journal} {2D Mater.}\ }\textbf {\bibinfo {volume} {6}},\ \bibinfo {pages}
  {015024} (\bibinfo {year} {2018})}\BibitemShut {NoStop}%
\bibitem [{\citenamefont {Yu}\ \emph {et~al.}(2022)\citenamefont {Yu},
  \citenamefont {Kutana},\ and\ \citenamefont {Yakobson}}]{yu_2022}%
  \BibitemOpen
  \bibfield  {author} {\bibinfo {author} {\bibfnamefont {H.}~\bibnamefont
  {Yu}}, \bibinfo {author} {\bibfnamefont {A.}~\bibnamefont {Kutana}},\ and\
  \bibinfo {author} {\bibfnamefont {B.~I.}\ \bibnamefont {Yakobson}},\
  }\bibfield  {title} {\bibinfo {title} {{Electron Optics and Valley Hall
  Effect of Undulated Graphene}},\ }\href
  {https://doi.org/10.1021/acs.nanolett.2c00103} {\bibfield  {journal}
  {\bibinfo  {journal} {Nano Lett.}\ }\textbf {\bibinfo {volume} {22}},\
  \bibinfo {pages} {2934} (\bibinfo {year} {2022})}\BibitemShut {NoStop}%
\bibitem [{\citenamefont {Franklin}\ \emph {et~al.}(2012)\citenamefont
  {Franklin}, \citenamefont {Luisier}, \citenamefont {Han}, \citenamefont
  {Tulevski}, \citenamefont {Breslin}, \citenamefont {Gignac}, \citenamefont
  {Lundstrom},\ and\ \citenamefont {Haensch}}]{franklin_2012}%
  \BibitemOpen
  \bibfield  {author} {\bibinfo {author} {\bibfnamefont {A.~D.}\ \bibnamefont
  {Franklin}}, \bibinfo {author} {\bibfnamefont {M.}~\bibnamefont {Luisier}},
  \bibinfo {author} {\bibfnamefont {S.-J.}\ \bibnamefont {Han}}, \bibinfo
  {author} {\bibfnamefont {G.}~\bibnamefont {Tulevski}}, \bibinfo {author}
  {\bibfnamefont {C.~M.}\ \bibnamefont {Breslin}}, \bibinfo {author}
  {\bibfnamefont {L.}~\bibnamefont {Gignac}}, \bibinfo {author} {\bibfnamefont
  {M.~S.}\ \bibnamefont {Lundstrom}},\ and\ \bibinfo {author} {\bibfnamefont
  {W.}~\bibnamefont {Haensch}},\ }\bibfield  {title} {\bibinfo {title} {{Sub-10
  nm Carbon Nanotube Transistor}},\ }\href {https://doi.org/10.1021/nl203701g}
  {\bibfield  {journal} {\bibinfo  {journal} {Nano Lett.}\ }\textbf {\bibinfo
  {volume} {12}},\ \bibinfo {pages} {758} (\bibinfo {year} {2012})}\BibitemShut
  {NoStop}%
\bibitem [{\citenamefont {Shulaker}\ \emph {et~al.}(2013)\citenamefont
  {Shulaker}, \citenamefont {Hills}, \citenamefont {Patil}, \citenamefont
  {Wei}, \citenamefont {Chen}, \citenamefont {Wong},\ and\ \citenamefont
  {Mitra}}]{Shulaker2013}%
  \BibitemOpen
  \bibfield  {author} {\bibinfo {author} {\bibfnamefont {M.~M.}\ \bibnamefont
  {Shulaker}}, \bibinfo {author} {\bibfnamefont {G.}~\bibnamefont {Hills}},
  \bibinfo {author} {\bibfnamefont {N.}~\bibnamefont {Patil}}, \bibinfo
  {author} {\bibfnamefont {H.}~\bibnamefont {Wei}}, \bibinfo {author}
  {\bibfnamefont {H.-Y.}\ \bibnamefont {Chen}}, \bibinfo {author}
  {\bibfnamefont {H.~S.~P.}\ \bibnamefont {Wong}},\ and\ \bibinfo {author}
  {\bibfnamefont {S.}~\bibnamefont {Mitra}},\ }\bibfield  {title} {\bibinfo
  {title} {Carbon nanotube computer},\ }\href
  {https://doi.org/10.1038/nature12502} {\bibfield  {journal} {\bibinfo
  {journal} {Nature}\ }\textbf {\bibinfo {volume} {501}},\ \bibinfo {pages}
  {526} (\bibinfo {year} {2013})}\BibitemShut {NoStop}%
\bibitem [{\citenamefont {Tulevski}\ \emph {et~al.}(2014)\citenamefont
  {Tulevski}, \citenamefont {Franklin}, \citenamefont {Frank}, \citenamefont
  {Lobez}, \citenamefont {Cao}, \citenamefont {Park}, \citenamefont {Afzali},
  \citenamefont {Han}, \citenamefont {Hannon},\ and\ \citenamefont
  {Haensch}}]{tulevski_2014}%
  \BibitemOpen
  \bibfield  {author} {\bibinfo {author} {\bibfnamefont {G.~S.}\ \bibnamefont
  {Tulevski}}, \bibinfo {author} {\bibfnamefont {A.~D.}\ \bibnamefont
  {Franklin}}, \bibinfo {author} {\bibfnamefont {D.}~\bibnamefont {Frank}},
  \bibinfo {author} {\bibfnamefont {J.~M.}\ \bibnamefont {Lobez}}, \bibinfo
  {author} {\bibfnamefont {Q.}~\bibnamefont {Cao}}, \bibinfo {author}
  {\bibfnamefont {H.}~\bibnamefont {Park}}, \bibinfo {author} {\bibfnamefont
  {A.}~\bibnamefont {Afzali}}, \bibinfo {author} {\bibfnamefont {S.-J.}\
  \bibnamefont {Han}}, \bibinfo {author} {\bibfnamefont {J.~B.}\ \bibnamefont
  {Hannon}},\ and\ \bibinfo {author} {\bibfnamefont {W.}~\bibnamefont
  {Haensch}},\ }\bibfield  {title} {\bibinfo {title} {{Toward High-Performance
  Digital Logic Technology with Carbon Nanotubes}},\ }\href
  {https://doi.org/10.1021/nn503627h} {\bibfield  {journal} {\bibinfo
  {journal} {ACS Nano}\ }\textbf {\bibinfo {volume} {8}},\ \bibinfo {pages}
  {8730} (\bibinfo {year} {2014})}\BibitemShut {NoStop}%
\bibitem [{\citenamefont {Hills}\ \emph {et~al.}(2019)\citenamefont {Hills},
  \citenamefont {Lau}, \citenamefont {Wright}, \citenamefont {Fuller},
  \citenamefont {Bishop}, \citenamefont {Srimani}, \citenamefont {Kanhaiya},
  \citenamefont {Ho}, \citenamefont {Amer}, \citenamefont {Stein},
  \citenamefont {Murphy}, \citenamefont {Arvind}, \citenamefont
  {Chandrakasan},\ and\ \citenamefont {Shulaker}}]{Hills2019}%
  \BibitemOpen
  \bibfield  {author} {\bibinfo {author} {\bibfnamefont {G.}~\bibnamefont
  {Hills}}, \bibinfo {author} {\bibfnamefont {C.}~\bibnamefont {Lau}}, \bibinfo
  {author} {\bibfnamefont {A.}~\bibnamefont {Wright}}, \bibinfo {author}
  {\bibfnamefont {S.}~\bibnamefont {Fuller}}, \bibinfo {author} {\bibfnamefont
  {M.~D.}\ \bibnamefont {Bishop}}, \bibinfo {author} {\bibfnamefont
  {T.}~\bibnamefont {Srimani}}, \bibinfo {author} {\bibfnamefont
  {P.}~\bibnamefont {Kanhaiya}}, \bibinfo {author} {\bibfnamefont
  {R.}~\bibnamefont {Ho}}, \bibinfo {author} {\bibfnamefont {A.}~\bibnamefont
  {Amer}}, \bibinfo {author} {\bibfnamefont {Y.}~\bibnamefont {Stein}},
  \bibinfo {author} {\bibfnamefont {D.}~\bibnamefont {Murphy}}, \bibinfo
  {author} {\bibnamefont {Arvind}}, \bibinfo {author} {\bibfnamefont
  {A.}~\bibnamefont {Chandrakasan}},\ and\ \bibinfo {author} {\bibfnamefont
  {M.~M.}\ \bibnamefont {Shulaker}},\ }\bibfield  {title} {\bibinfo {title}
  {Modern microprocessor built from complementary carbon nanotube
  transistors},\ }\href {https://doi.org/10.1038/s41586-019-1493-8} {\bibfield
  {journal} {\bibinfo  {journal} {Nature}\ }\textbf {\bibinfo {volume} {572}},\
  \bibinfo {pages} {595} (\bibinfo {year} {2019})}\BibitemShut {NoStop}%
\bibitem [{\citenamefont {Brand}\ \emph {et~al.}(2008)\citenamefont {Brand},
  \citenamefont {Fedder}, \citenamefont {Korvink}, \citenamefont {Tabata},\
  and\ \citenamefont {Hierold}}]{Brand2008}%
  \BibitemOpen
  \bibfield  {author} {\bibinfo {author} {\bibfnamefont {O.}~\bibnamefont
  {Brand}}, \bibinfo {author} {\bibfnamefont {G.~K.}\ \bibnamefont {Fedder}},
  \bibinfo {author} {\bibfnamefont {J.~G.}\ \bibnamefont {Korvink}}, \bibinfo
  {author} {\bibfnamefont {O.}~\bibnamefont {Tabata}},\ and\ \bibinfo {author}
  {\bibfnamefont {C.}~\bibnamefont {Hierold}},\ }\href
  {https://doi.org/10.1002/9783527622597} {\emph {\bibinfo {title} {Carbon
  Nanotube Devices: Properties, Modeling, Integration and Applications}}}\
  (\bibinfo  {publisher} {Wiley-VCH, Weinheim},\ \bibinfo {year}
  {2008})\BibitemShut {NoStop}%
\bibitem [{\citenamefont {Todri-Sanial}\ \emph {et~al.}(2016)\citenamefont
  {Todri-Sanial}, \citenamefont {Dijon},\ and\ \citenamefont
  {Maffucci}}]{Todri2017}%
  \BibitemOpen
  \bibfield  {author} {\bibinfo {author} {\bibfnamefont {A.}~\bibnamefont
  {Todri-Sanial}}, \bibinfo {author} {\bibfnamefont {J.}~\bibnamefont
  {Dijon}},\ and\ \bibinfo {author} {\bibfnamefont {A.}~\bibnamefont
  {Maffucci}},\ }\href {https://doi.org/10.1007/978-3-319-29746-0} {\emph
  {\bibinfo {title} {Carbon Nanotubes for Interconnects}}}\ (\bibinfo
  {publisher} {Springer, Cham},\ \bibinfo {year} {2016})\BibitemShut {NoStop}%
\bibitem [{\citenamefont {Hierold}\ \emph {et~al.}(2007)\citenamefont
  {Hierold}, \citenamefont {Jungen}, \citenamefont {Stampfer},\ and\
  \citenamefont {Helbling}}]{hierold_2007}%
  \BibitemOpen
  \bibfield  {author} {\bibinfo {author} {\bibfnamefont {C.}~\bibnamefont
  {Hierold}}, \bibinfo {author} {\bibfnamefont {A.}~\bibnamefont {Jungen}},
  \bibinfo {author} {\bibfnamefont {C.}~\bibnamefont {Stampfer}},\ and\
  \bibinfo {author} {\bibfnamefont {T.}~\bibnamefont {Helbling}},\ }\bibfield
  {title} {\bibinfo {title} {Nano electromechanical sensors based on carbon
  nanotubes},\ }\href
  {https://doi.org/https://doi.org/10.1016/j.sna.2007.02.007} {\bibfield
  {journal} {\bibinfo  {journal} {Sens. Actuators A Phys.}\ }\textbf {\bibinfo
  {volume} {136}},\ \bibinfo {pages} {51} (\bibinfo {year} {2007})}\BibitemShut
  {NoStop}%
\bibitem [{\citenamefont {Schroeder}\ \emph {et~al.}(2019)\citenamefont
  {Schroeder}, \citenamefont {Savagatrup}, \citenamefont {He}, \citenamefont
  {Lin},\ and\ \citenamefont {Swager}}]{Schroeder2018}%
  \BibitemOpen
  \bibfield  {author} {\bibinfo {author} {\bibfnamefont {V.}~\bibnamefont
  {Schroeder}}, \bibinfo {author} {\bibfnamefont {S.}~\bibnamefont
  {Savagatrup}}, \bibinfo {author} {\bibfnamefont {M.}~\bibnamefont {He}},
  \bibinfo {author} {\bibfnamefont {S.}~\bibnamefont {Lin}},\ and\ \bibinfo
  {author} {\bibfnamefont {T.~M.}\ \bibnamefont {Swager}},\ }\bibfield  {title}
  {\bibinfo {title} {{Carbon Nanotube Chemical Sensors}},\ }\href
  {https://doi.org/10.1021/acs.chemrev.8b00340} {\bibfield  {journal} {\bibinfo
   {journal} {Chem. Rev.}\ }\textbf {\bibinfo {volume} {119}},\ \bibinfo
  {pages} {599} (\bibinfo {year} {2019})}\BibitemShut {NoStop}%
\bibitem [{\citenamefont {Carrillo-Bastos}\ \emph {et~al.}(2016)\citenamefont
  {Carrillo-Bastos}, \citenamefont {Le\'on}, \citenamefont {Faria},
  \citenamefont {Latg\'e}, \citenamefont {Andrei},\ and\ \citenamefont
  {Sandler}}]{Carrillo-Bastos2016}%
  \BibitemOpen
  \bibfield  {author} {\bibinfo {author} {\bibfnamefont {R.}~\bibnamefont
  {Carrillo-Bastos}}, \bibinfo {author} {\bibfnamefont {C.}~\bibnamefont
  {Le\'on}}, \bibinfo {author} {\bibfnamefont {D.}~\bibnamefont {Faria}},
  \bibinfo {author} {\bibfnamefont {A.}~\bibnamefont {Latg\'e}}, \bibinfo
  {author} {\bibfnamefont {E.~Y.}\ \bibnamefont {Andrei}},\ and\ \bibinfo
  {author} {\bibfnamefont {N.}~\bibnamefont {Sandler}},\ }\bibfield  {title}
  {\bibinfo {title} {Strained fold-assisted transport in graphene systems},\
  }\href {https://doi.org/10.1103/PhysRevB.94.125422} {\bibfield  {journal}
  {\bibinfo  {journal} {Phys. Rev. B}\ }\textbf {\bibinfo {volume} {94}},\
  \bibinfo {pages} {125422} (\bibinfo {year} {2016})}\BibitemShut {NoStop}%
\bibitem [{\citenamefont {Pereira}\ \emph {et~al.}(2009)\citenamefont
  {Pereira}, \citenamefont {Castro~Neto},\ and\ \citenamefont
  {Peres}}]{Pereira2009}%
  \BibitemOpen
  \bibfield  {author} {\bibinfo {author} {\bibfnamefont {V.~M.}\ \bibnamefont
  {Pereira}}, \bibinfo {author} {\bibfnamefont {A.~H.}\ \bibnamefont
  {Castro~Neto}},\ and\ \bibinfo {author} {\bibfnamefont {N.~M.~R.}\
  \bibnamefont {Peres}},\ }\bibfield  {title} {\bibinfo {title} {Tight-binding
  approach to uniaxial strain in graphene},\ }\href
  {https://doi.org/10.1103/PhysRevB.80.045401} {\bibfield  {journal} {\bibinfo
  {journal} {Phys. Rev. B}\ }\textbf {\bibinfo {volume} {80}},\ \bibinfo
  {pages} {045401} (\bibinfo {year} {2009})}\BibitemShut {NoStop}%
\bibitem [{\citenamefont {Ribeiro}\ \emph {et~al.}(2009)\citenamefont
  {Ribeiro}, \citenamefont {Pereira}, \citenamefont {Peres}, \citenamefont
  {Briddon},\ and\ \citenamefont {Neto}}]{Ribeiro2009}%
  \BibitemOpen
  \bibfield  {author} {\bibinfo {author} {\bibfnamefont {R.~M.}\ \bibnamefont
  {Ribeiro}}, \bibinfo {author} {\bibfnamefont {V.~M.}\ \bibnamefont
  {Pereira}}, \bibinfo {author} {\bibfnamefont {N.~M.~R.}\ \bibnamefont
  {Peres}}, \bibinfo {author} {\bibfnamefont {P.~R.}\ \bibnamefont {Briddon}},\
  and\ \bibinfo {author} {\bibfnamefont {A.~H.~C.}\ \bibnamefont {Neto}},\
  }\bibfield  {title} {\bibinfo {title} {Strained graphene: {t}ight-binding and
  density functional calculations},\ }\href
  {https://doi.org/10.1088/1367-2630/11/11/115002} {\bibfield  {journal}
  {\bibinfo  {journal} {New J. Phys.}\ }\textbf {\bibinfo {volume} {11}},\
  \bibinfo {pages} {115002} (\bibinfo {year} {2009})}\BibitemShut {NoStop}%
\bibitem [{\citenamefont {Kleiner}\ and\ \citenamefont
  {Eggert}(2001{\natexlab{a}})}]{Kleiner+Egger1}%
  \BibitemOpen
  \bibfield  {author} {\bibinfo {author} {\bibfnamefont {A.}~\bibnamefont
  {Kleiner}}\ and\ \bibinfo {author} {\bibfnamefont {S.}~\bibnamefont
  {Eggert}},\ }\bibfield  {title} {\bibinfo {title} {Band gaps of primary
  metallic carbon nanotubes},\ }\href
  {https://doi.org/10.1103/PhysRevB.63.073408} {\bibfield  {journal} {\bibinfo
  {journal} {Phys. Rev. B}\ }\textbf {\bibinfo {volume} {63}},\ \bibinfo
  {pages} {073408} (\bibinfo {year} {2001}{\natexlab{a}})}\BibitemShut
  {NoStop}%
\bibitem [{\citenamefont {Kleiner}\ and\ \citenamefont
  {Eggert}(2001{\natexlab{b}})}]{Kleiner+Egger2}%
  \BibitemOpen
  \bibfield  {author} {\bibinfo {author} {\bibfnamefont {A.}~\bibnamefont
  {Kleiner}}\ and\ \bibinfo {author} {\bibfnamefont {S.}~\bibnamefont
  {Eggert}},\ }\bibfield  {title} {\bibinfo {title} {{Curvature, hybridization,
  and STM images of carbon nanotubes}},\ }\href
  {https://doi.org/10.1103/PhysRevB.64.113402} {\bibfield  {journal} {\bibinfo
  {journal} {Phys. Rev. B}\ }\textbf {\bibinfo {volume} {64}},\ \bibinfo
  {pages} {113402} (\bibinfo {year} {2001}{\natexlab{b}})}\BibitemShut
  {NoStop}%
\bibitem [{\citenamefont {Gmitra}\ \emph {et~al.}(2009)\citenamefont {Gmitra},
  \citenamefont {Konschuh}, \citenamefont {Ertler}, \citenamefont
  {Ambrosch-Draxl},\ and\ \citenamefont {Fabian}}]{gmitra2009band}%
  \BibitemOpen
  \bibfield  {author} {\bibinfo {author} {\bibfnamefont {M.}~\bibnamefont
  {Gmitra}}, \bibinfo {author} {\bibfnamefont {S.}~\bibnamefont {Konschuh}},
  \bibinfo {author} {\bibfnamefont {C.}~\bibnamefont {Ertler}}, \bibinfo
  {author} {\bibfnamefont {C.}~\bibnamefont {Ambrosch-Draxl}},\ and\ \bibinfo
  {author} {\bibfnamefont {J.}~\bibnamefont {Fabian}},\ }\bibfield  {title}
  {\bibinfo {title} {Band-structure topologies of graphene: Spin-orbit coupling
  effects from first principles},\ }\href
  {https://doi.org/10.1103/PhysRevB.80.235431} {\bibfield  {journal} {\bibinfo
  {journal} {Phys. Rev. B}\ }\textbf {\bibinfo {volume} {80}},\ \bibinfo
  {pages} {235431} (\bibinfo {year} {2009})}\BibitemShut {NoStop}%
\bibitem [{\citenamefont {Huertas-Hernando}\ \emph {et~al.}(2006)\citenamefont
  {Huertas-Hernando}, \citenamefont {Guinea},\ and\ \citenamefont
  {Brataas}}]{huertas_2006}%
  \BibitemOpen
  \bibfield  {author} {\bibinfo {author} {\bibfnamefont {D.}~\bibnamefont
  {Huertas-Hernando}}, \bibinfo {author} {\bibfnamefont {F.}~\bibnamefont
  {Guinea}},\ and\ \bibinfo {author} {\bibfnamefont {A.}~\bibnamefont
  {Brataas}},\ }\bibfield  {title} {\bibinfo {title} {Spin-orbit coupling in
  curved graphene, fullerenes, nanotubes, and nanotube caps},\ }\href
  {https://doi.org/10.1103/PhysRevB.74.155426} {\bibfield  {journal} {\bibinfo
  {journal} {Phys. Rev. B}\ }\textbf {\bibinfo {volume} {74}},\ \bibinfo
  {pages} {155426} (\bibinfo {year} {2006})}\BibitemShut {NoStop}%
\bibitem [{\citenamefont {Jeong}\ and\ \citenamefont {Lee}(2009)}]{jeong_2009}%
  \BibitemOpen
  \bibfield  {author} {\bibinfo {author} {\bibfnamefont {J.-S.}\ \bibnamefont
  {Jeong}}\ and\ \bibinfo {author} {\bibfnamefont {H.-W.}\ \bibnamefont
  {Lee}},\ }\bibfield  {title} {\bibinfo {title} {Curvature-enhanced spin-orbit
  coupling in a carbon nanotube},\ }\href
  {https://doi.org/10.1103/PhysRevB.80.075409} {\bibfield  {journal} {\bibinfo
  {journal} {Phys. Rev. B}\ }\textbf {\bibinfo {volume} {80}},\ \bibinfo
  {pages} {075409} (\bibinfo {year} {2009})}\BibitemShut {NoStop}%
\bibitem [{\citenamefont {Izumida}\ \emph {et~al.}(2009)\citenamefont
  {Izumida}, \citenamefont {Sato},\ and\ \citenamefont {Saito}}]{izumida_2009}%
  \BibitemOpen
  \bibfield  {author} {\bibinfo {author} {\bibfnamefont {W.}~\bibnamefont
  {Izumida}}, \bibinfo {author} {\bibfnamefont {K.}~\bibnamefont {Sato}},\ and\
  \bibinfo {author} {\bibfnamefont {R.}~\bibnamefont {Saito}},\ }\bibfield
  {title} {\bibinfo {title} {{Spin--Orbit Interaction in Single Wall Carbon
  Nanotubes: Symmetry Adapted Tight-Binding Calculation and Effective Model
  Analysis}},\ }\href {https://doi.org/10.1143/JPSJ.78.074707} {\bibfield
  {journal} {\bibinfo  {journal} {J. Phys. Soc. Japan}\ }\textbf {\bibinfo
  {volume} {78}},\ \bibinfo {pages} {074707} (\bibinfo {year}
  {2009})}\BibitemShut {NoStop}%
\bibitem [{\citenamefont {Klinovaja}\ \emph {et~al.}(2011)\citenamefont
  {Klinovaja}, \citenamefont {Schmidt}, \citenamefont {Braunecker},\ and\
  \citenamefont {Loss}}]{klinovaja_2011}%
  \BibitemOpen
  \bibfield  {author} {\bibinfo {author} {\bibfnamefont {J.}~\bibnamefont
  {Klinovaja}}, \bibinfo {author} {\bibfnamefont {M.~J.}\ \bibnamefont
  {Schmidt}}, \bibinfo {author} {\bibfnamefont {B.}~\bibnamefont
  {Braunecker}},\ and\ \bibinfo {author} {\bibfnamefont {D.}~\bibnamefont
  {Loss}},\ }\bibfield  {title} {\bibinfo {title} {Carbon nanotubes in electric
  and magnetic fields},\ }\href {https://doi.org/10.1103/PhysRevB.84.085452}
  {\bibfield  {journal} {\bibinfo  {journal} {Phys. Rev. B}\ }\textbf {\bibinfo
  {volume} {84}},\ \bibinfo {pages} {085452} (\bibinfo {year}
  {2011})}\BibitemShut {NoStop}%
\bibitem [{\citenamefont {Nakada}\ \emph {et~al.}(1996)\citenamefont {Nakada},
  \citenamefont {Fujita}, \citenamefont {Dresselhaus},\ and\ \citenamefont
  {Dresselhaus}}]{nakada_1996}%
  \BibitemOpen
  \bibfield  {author} {\bibinfo {author} {\bibfnamefont {K.}~\bibnamefont
  {Nakada}}, \bibinfo {author} {\bibfnamefont {M.}~\bibnamefont {Fujita}},
  \bibinfo {author} {\bibfnamefont {G.}~\bibnamefont {Dresselhaus}},\ and\
  \bibinfo {author} {\bibfnamefont {M.~S.}\ \bibnamefont {Dresselhaus}},\
  }\bibfield  {title} {\bibinfo {title} {{Edge state in graphene ribbons:
  Nanometer size effect and edge shape dependence}},\ }\href
  {https://doi.org/10.1103/PhysRevB.54.17954} {\bibfield  {journal} {\bibinfo
  {journal} {Phys. Rev. B}\ }\textbf {\bibinfo {volume} {54}},\ \bibinfo
  {pages} {17954} (\bibinfo {year} {1996})}\BibitemShut {NoStop}%
\bibitem [{\citenamefont {Sasaki}\ \emph {et~al.}(2005)\citenamefont {Sasaki},
  \citenamefont {Murakami}, \citenamefont {Saito},\ and\ \citenamefont
  {Kawazoe}}]{sasaki_2005}%
  \BibitemOpen
  \bibfield  {author} {\bibinfo {author} {\bibfnamefont {K.}~\bibnamefont
  {Sasaki}}, \bibinfo {author} {\bibfnamefont {S.}~\bibnamefont {Murakami}},
  \bibinfo {author} {\bibfnamefont {R.}~\bibnamefont {Saito}},\ and\ \bibinfo
  {author} {\bibfnamefont {Y.}~\bibnamefont {Kawazoe}},\ }\bibfield  {title}
  {\bibinfo {title} {{Controlling edge states of zigzag carbon nanotubes by the
  Aharonov-Bohm flux}},\ }\href {https://doi.org/10.1103/PhysRevB.71.195401}
  {\bibfield  {journal} {\bibinfo  {journal} {Phys. Rev. B}\ }\textbf {\bibinfo
  {volume} {71}},\ \bibinfo {pages} {195401} (\bibinfo {year}
  {2005})}\BibitemShut {NoStop}%
\bibitem [{\citenamefont {Choi}\ \emph {et~al.}(2010)\citenamefont {Choi},
  \citenamefont {Jhi},\ and\ \citenamefont {Son}}]{seon-myeong_2010}%
  \BibitemOpen
  \bibfield  {author} {\bibinfo {author} {\bibfnamefont {S.-M.}\ \bibnamefont
  {Choi}}, \bibinfo {author} {\bibfnamefont {S.-H.}\ \bibnamefont {Jhi}},\ and\
  \bibinfo {author} {\bibfnamefont {Y.-W.}\ \bibnamefont {Son}},\ }\bibfield
  {title} {\bibinfo {title} {{Effects of strain on electronic properties of
  graphene}},\ }\href {https://doi.org/10.1103/PhysRevB.81.081407} {\bibfield
  {journal} {\bibinfo  {journal} {Phys. Rev. B}\ }\textbf {\bibinfo {volume}
  {81}},\ \bibinfo {pages} {081407(R)} (\bibinfo {year} {2010})}\BibitemShut
  {NoStop}%
\bibitem [{\citenamefont {Levy}\ \emph {et~al.}(2010)\citenamefont {Levy},
  \citenamefont {Burke}, \citenamefont {Meaker}, \citenamefont {Panlasigui},
  \citenamefont {Zettl}, \citenamefont {Guinea}, \citenamefont {Neto},\ and\
  \citenamefont {Crommie}}]{levy_2010}%
  \BibitemOpen
  \bibfield  {author} {\bibinfo {author} {\bibfnamefont {N.}~\bibnamefont
  {Levy}}, \bibinfo {author} {\bibfnamefont {S.~A.}\ \bibnamefont {Burke}},
  \bibinfo {author} {\bibfnamefont {K.~L.}\ \bibnamefont {Meaker}}, \bibinfo
  {author} {\bibfnamefont {M.}~\bibnamefont {Panlasigui}}, \bibinfo {author}
  {\bibfnamefont {A.}~\bibnamefont {Zettl}}, \bibinfo {author} {\bibfnamefont
  {F.}~\bibnamefont {Guinea}}, \bibinfo {author} {\bibfnamefont {A.~H.~C.}\
  \bibnamefont {Neto}},\ and\ \bibinfo {author} {\bibfnamefont {M.~F.}\
  \bibnamefont {Crommie}},\ }\bibfield  {title} {\bibinfo {title}
  {{Strain-Induced Pseudo-Magnetic Fields Greater Than 300 Tesla in Graphene
  Nanobubbles}},\ }\href {https://doi.org/10.1126/science.1191700} {\bibfield
  {journal} {\bibinfo  {journal} {Science}\ }\textbf {\bibinfo {volume}
  {329}},\ \bibinfo {pages} {544} (\bibinfo {year} {2010})}\BibitemShut
  {NoStop}%
\bibitem [{\citenamefont {Kim}\ \emph {et~al.}(2011)\citenamefont {Kim},
  \citenamefont {Blanter},\ and\ \citenamefont {Ahn}}]{kim_2011}%
  \BibitemOpen
  \bibfield  {author} {\bibinfo {author} {\bibfnamefont {K.-J.}\ \bibnamefont
  {Kim}}, \bibinfo {author} {\bibfnamefont {Y.~M.}\ \bibnamefont {Blanter}},\
  and\ \bibinfo {author} {\bibfnamefont {K.-H.}\ \bibnamefont {Ahn}},\
  }\bibfield  {title} {\bibinfo {title} {Interplay between real and
  pseudomagnetic field in graphene with strain},\ }\href
  {https://doi.org/10.1103/PhysRevB.84.081401} {\bibfield  {journal} {\bibinfo
  {journal} {Phys. Rev. B}\ }\textbf {\bibinfo {volume} {84}},\ \bibinfo
  {pages} {081401(R)} (\bibinfo {year} {2011})}\BibitemShut {NoStop}%
\bibitem [{\citenamefont {de~Juan}\ \emph {et~al.}(2012)\citenamefont
  {de~Juan}, \citenamefont {Sturla},\ and\ \citenamefont
  {Vozmediano}}]{dejuan_2012}%
  \BibitemOpen
  \bibfield  {author} {\bibinfo {author} {\bibfnamefont {F.}~\bibnamefont
  {de~Juan}}, \bibinfo {author} {\bibfnamefont {M.}~\bibnamefont {Sturla}},\
  and\ \bibinfo {author} {\bibfnamefont {M.~A.~H.}\ \bibnamefont
  {Vozmediano}},\ }\bibfield  {title} {\bibinfo {title} {{Space Dependent Fermi
  Velocity in Strained Graphene}},\ }\href
  {https://doi.org/10.1103/PhysRevLett.108.227205} {\bibfield  {journal}
  {\bibinfo  {journal} {Phys. Rev. Lett.}\ }\textbf {\bibinfo {volume} {108}},\
  \bibinfo {pages} {227205} (\bibinfo {year} {2012})}\BibitemShut {NoStop}%
\bibitem [{\citenamefont {Kozlovsky}\ \emph {et~al.}(2020)\citenamefont
  {Kozlovsky}, \citenamefont {Graf}, \citenamefont {Kochan}, \citenamefont
  {Richter},\ and\ \citenamefont {Gorini}}]{kozlovsky_2020}%
  \BibitemOpen
  \bibfield  {author} {\bibinfo {author} {\bibfnamefont {R.}~\bibnamefont
  {Kozlovsky}}, \bibinfo {author} {\bibfnamefont {A.}~\bibnamefont {Graf}},
  \bibinfo {author} {\bibfnamefont {D.}~\bibnamefont {Kochan}}, \bibinfo
  {author} {\bibfnamefont {K.}~\bibnamefont {Richter}},\ and\ \bibinfo {author}
  {\bibfnamefont {C.}~\bibnamefont {Gorini}},\ }\bibfield  {title} {\bibinfo
  {title} {{Magnetoconductance, Quantum Hall Effect, and Coulomb Blockade in
  Topological Insulator Nanocones}},\ }\href
  {https://doi.org/10.1103/PhysRevLett.124.126804} {\bibfield  {journal}
  {\bibinfo  {journal} {Phys. Rev. Lett.}\ }\textbf {\bibinfo {volume} {124}},\
  \bibinfo {pages} {126804} (\bibinfo {year} {2020})}\BibitemShut {NoStop}%
\bibitem [{\citenamefont {Abramowitz}\ and\ \citenamefont
  {Stegun}(1948)}]{abramowitz_1988}%
  \BibitemOpen
  \bibfield  {author} {\bibinfo {author} {\bibfnamefont {M.}~\bibnamefont
  {Abramowitz}}\ and\ \bibinfo {author} {\bibfnamefont {I.~A.}\ \bibnamefont
  {Stegun}},\ }\href@noop {} {\emph {\bibinfo {title} {Handbook of Mathematical
  Functions With Formulas, Graphs, and Mathematical Tables}}},\ Vol.~\bibinfo
  {volume} {55}\ (\bibinfo  {publisher} {US Government Printing Office},\
  \bibinfo {year} {1948})\BibitemShut {NoStop}%
\bibitem [{\citenamefont {Caroli}\ \emph {et~al.}(1971)\citenamefont {Caroli},
  \citenamefont {Combescot}, \citenamefont {Nozieres},\ and\ \citenamefont
  {Saint-James}}]{caroli_1971}%
  \BibitemOpen
  \bibfield  {author} {\bibinfo {author} {\bibfnamefont {C.}~\bibnamefont
  {Caroli}}, \bibinfo {author} {\bibfnamefont {R.}~\bibnamefont {Combescot}},
  \bibinfo {author} {\bibfnamefont {P.}~\bibnamefont {Nozieres}},\ and\
  \bibinfo {author} {\bibfnamefont {D.}~\bibnamefont {Saint-James}},\
  }\bibfield  {title} {\bibinfo {title} {Direct calculation of the tunneling
  current},\ }\href {https://doi.org/10.1088/0022-3719/4/8/018} {\bibfield
  {journal} {\bibinfo  {journal} {J. Solid State Phys.}\ }\textbf {\bibinfo
  {volume} {4}},\ \bibinfo {pages} {916} (\bibinfo {year} {1971})}\BibitemShut
  {NoStop}%
\bibitem [{\citenamefont {Cresti}\ \emph {et~al.}(2003)\citenamefont {Cresti},
  \citenamefont {Farchioni}, \citenamefont {Grosso},\ and\ \citenamefont
  {Parravicini}}]{cresti_2003}%
  \BibitemOpen
  \bibfield  {author} {\bibinfo {author} {\bibfnamefont {A.}~\bibnamefont
  {Cresti}}, \bibinfo {author} {\bibfnamefont {R.}~\bibnamefont {Farchioni}},
  \bibinfo {author} {\bibfnamefont {G.}~\bibnamefont {Grosso}},\ and\ \bibinfo
  {author} {\bibfnamefont {G.~P.}\ \bibnamefont {Parravicini}},\ }\bibfield
  {title} {\bibinfo {title} {Keldysh-{G}reen function formalism for current
  profiles in mesoscopic systems},\ }\href
  {https://doi.org/10.1103/PhysRevB.68.075306} {\bibfield  {journal} {\bibinfo
  {journal} {Phys. Rev. B}\ }\textbf {\bibinfo {volume} {68}},\ \bibinfo
  {pages} {075306} (\bibinfo {year} {2003})}\BibitemShut {NoStop}%
\bibitem [{\citenamefont {Zienert}\ \emph {et~al.}(2010)\citenamefont
  {Zienert}, \citenamefont {Schuster}, \citenamefont {Streiter},\ and\
  \citenamefont {Gessner}}]{zienert_2010}%
  \BibitemOpen
  \bibfield  {author} {\bibinfo {author} {\bibfnamefont {A.}~\bibnamefont
  {Zienert}}, \bibinfo {author} {\bibfnamefont {J.}~\bibnamefont {Schuster}},
  \bibinfo {author} {\bibfnamefont {R.}~\bibnamefont {Streiter}},\ and\
  \bibinfo {author} {\bibfnamefont {T.}~\bibnamefont {Gessner}},\ }\bibfield
  {title} {\bibinfo {title} {Transport in carbon nanotubes: Contact models and
  size effects},\ }\href
  {https://doi.org/https://doi.org/10.1002/pssb.201000178} {\bibfield
  {journal} {\bibinfo  {journal} {Phys. Status Solidi B}\ }\textbf {\bibinfo
  {volume} {247}},\ \bibinfo {pages} {3002} (\bibinfo {year}
  {2010})}\BibitemShut {NoStop}%
\bibitem [{\citenamefont {Lewenkopf}\ and\ \citenamefont
  {Mucciolo}(2013)}]{lewenkopf_2013}%
  \BibitemOpen
  \bibfield  {author} {\bibinfo {author} {\bibfnamefont {C.~H.}\ \bibnamefont
  {Lewenkopf}}\ and\ \bibinfo {author} {\bibfnamefont {E.~R.}\ \bibnamefont
  {Mucciolo}},\ }\bibfield  {title} {\bibinfo {title} {The recursive {G}reen's
  function method for graphene},\ }\href
  {https://doi.org/10.1007/s10825-013-0458-7} {\bibfield  {journal} {\bibinfo
  {journal} {J. Comput. Electron.}\ }\textbf {\bibinfo {volume} {12}},\
  \bibinfo {pages} {203} (\bibinfo {year} {2013})}\BibitemShut {NoStop}%
\bibitem [{\citenamefont {Settnes}\ \emph {et~al.}(2015)\citenamefont
  {Settnes}, \citenamefont {Power}, \citenamefont {Lin}, \citenamefont
  {Petersen},\ and\ \citenamefont {Jauho}}]{settnes_2015}%
  \BibitemOpen
  \bibfield  {author} {\bibinfo {author} {\bibfnamefont {M.}~\bibnamefont
  {Settnes}}, \bibinfo {author} {\bibfnamefont {S.~R.}\ \bibnamefont {Power}},
  \bibinfo {author} {\bibfnamefont {J.}~\bibnamefont {Lin}}, \bibinfo {author}
  {\bibfnamefont {D.~H.}\ \bibnamefont {Petersen}},\ and\ \bibinfo {author}
  {\bibfnamefont {A.-P.}\ \bibnamefont {Jauho}},\ }\bibfield  {title} {\bibinfo
  {title} {Patched {G}reen's function techniques for two-dimensional systems:
  Electronic behavior of bubbles and perforations in graphene},\ }\href
  {https://doi.org/10.1103/PhysRevB.91.125408} {\bibfield  {journal} {\bibinfo
  {journal} {Phys. Rev. B}\ }\textbf {\bibinfo {volume} {91}},\ \bibinfo
  {pages} {125408} (\bibinfo {year} {2015})}\BibitemShut {NoStop}%
\bibitem [{\citenamefont {Marga\ifmmode~\acute{n}\else \'{n}\fi{}ska}\ \emph
  {et~al.}(2011)\citenamefont {Marga\ifmmode~\acute{n}\else \'{n}\fi{}ska},
  \citenamefont {del Valle}, \citenamefont {Jhang}, \citenamefont {Strunk},\
  and\ \citenamefont {Grifoni}}]{marganska_2011}%
  \BibitemOpen
  \bibfield  {author} {\bibinfo {author} {\bibfnamefont {M.}~\bibnamefont
  {Marga\ifmmode~\acute{n}\else \'{n}\fi{}ska}}, \bibinfo {author}
  {\bibfnamefont {M.}~\bibnamefont {del Valle}}, \bibinfo {author}
  {\bibfnamefont {S.~H.}\ \bibnamefont {Jhang}}, \bibinfo {author}
  {\bibfnamefont {C.}~\bibnamefont {Strunk}},\ and\ \bibinfo {author}
  {\bibfnamefont {M.}~\bibnamefont {Grifoni}},\ }\bibfield  {title} {\bibinfo
  {title} {Localization induced by magnetic fields in carbon nanotubes},\
  }\href {https://doi.org/10.1103/PhysRevB.83.193407} {\bibfield  {journal}
  {\bibinfo  {journal} {Phys. Rev. B}\ }\textbf {\bibinfo {volume} {83}},\
  \bibinfo {pages} {193407} (\bibinfo {year} {2011})}\BibitemShut {NoStop}%
\bibitem [{\citenamefont {Teichert}\ \emph {et~al.}(2018)\citenamefont
  {Teichert}, \citenamefont {Wagner}, \citenamefont {Croy},\ and\ \citenamefont
  {Schuster}}]{teichert_2018}%
  \BibitemOpen
  \bibfield  {author} {\bibinfo {author} {\bibfnamefont {F.}~\bibnamefont
  {Teichert}}, \bibinfo {author} {\bibfnamefont {C.}~\bibnamefont {Wagner}},
  \bibinfo {author} {\bibfnamefont {A.}~\bibnamefont {Croy}},\ and\ \bibinfo
  {author} {\bibfnamefont {J.}~\bibnamefont {Schuster}},\ }\bibfield  {title}
  {\bibinfo {title} {Influence of defect-induced deformations on electron
  transport in carbon nanotubes},\ }\href
  {https://doi.org/10.1088/2399-6528/aaf08c} {\bibfield  {journal} {\bibinfo
  {journal} {J. Phys. Commun.}\ }\textbf {\bibinfo {volume} {2}},\ \bibinfo
  {pages} {115023} (\bibinfo {year} {2018})}\BibitemShut {NoStop}%
\bibitem [{\citenamefont {Wald}(1984)}]{wald2010general}%
  \BibitemOpen
  \bibfield  {author} {\bibinfo {author} {\bibfnamefont {R.~M.}\ \bibnamefont
  {Wald}},\ }\href {https://doi.org/10.7208/chicago/9780226870373.001.0001}
  {\emph {\bibinfo {title} {{General Relativity}}}}\ (\bibinfo  {publisher}
  {University of Chicago Press},\ \bibinfo {year} {1984})\BibitemShut {NoStop}%
\end{thebibliography}%

%%%%%%%%%%%%%%%%%%%%%%%%%%%%%%%%%%%%%%%%%%%%%%%%%%%%%%%%%%%%%%%%%%%%%%%%%%%%%%%%%%%%%%%%%%%%%%%%%%%%
%%%%%%%%%%%%%%%%%%%%%%%%%%%%%%%%%%%%%%%%%%%%%%%%%%%%%%%%%%%%%%%%%%%%%%%%%%%%%%%%%%%%%%%%%%%%%%%%%%%%
\end{document}